\documentclass[prd,nofootinbib,amsfonts,notitlepage]{revtex4-2}
\usepackage[utf8]{inputenc}
\usepackage[T1]{fontenc}
\usepackage{hyperref}
\usepackage{graphicx,bm,amsmath,color}
\usepackage[dvipsnames]{xcolor}
\usepackage{bbold}
\usepackage{wasysym}
\usepackage{verbatim}
\usepackage{amssymb}

\hypersetup{
colorlinks=true,         
linkcolor=Orange,          
citecolor=OliveGreen,           
urlcolor=Fuchsia   
}

\usepackage{epstopdf}
\usepackage{animate}
\usepackage{xmpmulti}
\usepackage{centernot}
\usepackage{multirow}
\usepackage{tikz}
\usepackage{graphicx}
\usepackage{float}
\usepackage{mathtools}
\usepackage{comment}

\makeatletter

\newcommand{\be}{\begin{equation}}
\newcommand{\ee}{\end{equation}}
\newcommand{\beq}{\begin{eqnarray}}
\newcommand{\eeq}{\end{eqnarray}}
\newcommand{\ba}{\begin{align}}
\newcommand{\ea}{\end{align}}

\begin{document}

\title{Relative Locality in curved spacetimes and event horizons}
\author{F. Mercati}
\affiliation{Departamento de Física, Universidad de Burgos, 09001 Burgos, Spain}
\email{fmercati@ubu.es}

\author{J.J. Relancio}
\affiliation{Departamento de Matemáticas y Computación, Universidad de Burgos, 09001 Burgos, Spain;\\ Centro de Astropartículas y Física de Altas Energías (CAPA),
Universidad de Zaragoza, Zaragoza 50009, Spain}
\email{jjrelancio@ubu.es}

\begin{abstract}
In the past decade, significant efforts have been devoted to the study of Relative Locality, which aims to generalize the kinematics of relativistic particles to a nonlocal framework by introducing a nontrivial geometry for momentum space. This paper builds upon a recent proposal to extend the theory to curved spacetimes and investigates the behavior of horizons in certain spacetimes with this nonlocality framework. Specifically, we examine whether nonlocality effects weaken or destroy the notion of horizon in these spacetimes. Our analysis indicates that, in the chosen models, the nonlocality effects do not disrupt the notion of horizon and that it remains as robust as it is in General Relativity.
\end{abstract}

\maketitle

\section{Introduction}

In a theory of  Quantum Gravity (QG), which aims to merge Quantum Field Theory (QFT) and General Relativity (GR), the classical notion of spacetime needs to be modified at small scales, leading to a ``quantum'' spacetime with novel properties that we are not yet able to fully model. This has been the subject of intense study over the years, with several competing theories proposing different models for a quantum spacetime. For example, in Loop Quantum Gravity, the structure of spacetime takes the form of a spin foam \citep{Sahlmann:2010zf,Dupuis:2012yw}, while Causal Set Theory and String Theory predict some form of nonlocality  \citep{Wallden:2013kka,Wallden:2010sh,Henson:2006kf,Mukhi:2011zz,Aharony:1999ks,Dienes:1996du,Belenchia:2014fda,Belenchia:2015ake}.

The concept of a quantum spacetime stands in contrast to Einstein's classical notion of spacetime \citep{Einstein1905}, which assumes that events can be described through the exchange of light signals. However, if a quantum spacetime exists, the propagation of massless particles could depend on their energy, and the classical description proposed by Einstein may not be sufficient when nonlocality effects arise. Therefore, a fundamental revision of the classical notion of spacetime seems inevitable in the context of QG.

The aforementioned theories are not yet fully satisfactory, as they lack well-defined and testable predictions. This motivates an alternative approach: instead of solely focusing on the search for a fundamental theory of QG, a ``top-down'' methodology can be attempted. This approach involves predicting the residual effects of a low-energy limit of a fundamental theory of QG, rather than starting from a fundamental theory. By doing so, it may be possible to give guidance to the experiments, in the search for a evidence of a quantum nature of spacetime, and propose novel physical phenomena that arise due to quantum gravity's effects on the low-energy regime. Such an approach  raises the question of how Lorentz invariance, a fundamental symmetry of GR (at the local level), might be affected. Lacking a fundamental theory to answer this question, the ``top-down'' approach requires one to make an assumption. Lorentz Invariance Violating (LIV) models have been studied extensively~\cite{Colladay:1998fq,Kostelecky:2008ts}, as well as theories which preserve the relativity principle, but realize it in a deformed way, \textit{i.e.}, Doubly Special Relativity (DSR) theories~\cite{Amelino-Camelia2002b,AmelinoCamelia:2001vy,AmelinoCamelia:2008qg}. The two approaches have very different phenomenologies~\cite{Addazi:2021xuf,AmelinoCamelia:2008qg,Amelino-Camelia:2011uwb,Carmona:2012un}, which suggest very different ways of testing their physical predictions.

In this paper, we are interested in the DSR scenario, which involves significantly harder conceptual challenges compared to the LIV one, as one is seeking to preserve the relativity principle while modifying the relativistic rules that connect different observers, and the consistency of the theory is always at risk. 
For example, after more than a decade of studies of DSR models, an incompatibility between the modified principle of relativity and the locality principle emerged, highlighting the necessity of weakening the notion of locality into a \textit{relative}, observer-dependent notion. Interactions appear local only to nearby observers, while distant observers might infer them as taking place in an extended region~\cite{AmelinoCamelia:2010qv}. This led to the development of the \textit{Relative Locality} (RL) proposal, \cite{AmelinoCamelia:2011bm,Amelino-Camelia:2011hjg,Gubitosi:2011hgc}, which reverses somewhat the roles of spacetime and the momentum space of a particle. In Special Relativity (SR), in fact, there is a formal duality (the so-called ``Born Reciprocity''~\cite{Born:1938}) between the coordinates and the relativistic momenta of a point particle, which both live on a spacetime with Minkowski geometry. GR breaks this duality, making spacetime curved and relating momentum space to the (flat) cotangent space at each point of spacetime. The basic idea of RL is to focus on a conjectured regime in which this duality is broken in the opposite way: momentum space is assumed to have a nontrivial geometry, and spacetime is related to its cotangent space, which is always flat. Then, the cotangent spaces at different points of momentum space have to be related through a nontrivial parallel transport, which encodes the curvature of momentum space, and is related to how different observers infer a spacetime out of the trajectories of point particles. This inference process is accurate only locally (near the origin of a reference frame). As one observes interactions further away, the vertices of these interactions become increasingly nonlocal in appearance.

One outstanding problem with RL theories is how to include in the picture spacetime curvature effects. This issue is not simply academic, as a phenomenology based on these theories requires one to look for effects that accumulated over cosmological distances, and the curvature of spacetime becomes important over these scales~\cite{Marcian2010InterplayBC,Amelino_Camelia_2012,Ballesteros2021InterplayBS}. 
In the present paper, we are interested in some conceptual issues that arise when considering a particular approach to this problem. A theory capable of describing general spacetime geometries within a RL model was developed in a series of papers~\cite{Relancio:2020zok,Relancio:2020mpa,Relancio:2020rys,Pfeifer:2021tas,Relancio:2021asx,Relancio:2022kpf} that built upon the geometric approach of Carmona et al. \cite{Carmona:2019fwf}. The theory reduces to the original proposal of~\cite{AmelinoCamelia:2011bm} in the flat limit.\footnote{This extension differs from the one considered in~\cite{Cianfrani:2014fia}. In that paper, an action with some nonlocal variables (defined by the space-time tetrad) is considered (differing from the approach of the aforementioned works in which the space-time coordinates are the canonical conjugated variables of the momentum), allowing them to generalize the RL action~\cite{AmelinoCamelia:2011bm} when a curvature in spacetime is present. As we will see, in our framework we are able to describe the RL principle in presence of a curvature on spacetime with the canonical variables, as it is done for the flat space-time case.}

Any theory that aspires to generalize GR and the locality principle into a consistent framework is likely to encounter significant conceptual challenges. In this paper, we explore one of the most glaring tension between the structure of GR and a weakened notion of locality: the fate of horizons and causal relations. One of the most striking predictions of GR, which sets it apart from SR, is the prediction of nontrivial causal relations. An observer in Minkoski space can receive a signal from anywhere in spacetime, as long as they can wait long enough. This is in stark contrast with what happens, for example, in Schwarzschild or de Sitter spacetime, where there are regions that can never communicate with certain observers~\cite{
hawking2023large,thorne2000gravitation}. These regions are separated from the observer by \textit{horizons} (an event horizon in the case of Schwarzschild, and a cosmological horizon in de Sitter). In GR, horizons are hypersurfaces, and are therefore ``sharp''. If the notion of locality is relaxed, one can expect the horizon of an observer to turn into a fuzzy, or even ill-defined concept. A superficial analysis of the structure of the theory of~\cite{Relancio:2020zok,Relancio:2020mpa,Relancio:2020rys,Pfeifer:2021tas,Relancio:2021asx,Relancio:2022kpf} reveals that, in principle, it allows for particles with sufficiently high-energy, that are created outside of an observer's horizon, to escape the horizon and reach the observer. This, of course, would have major consequences for many aspects of physics, notably for the physics of black holes. However, a detailed analysis of a concrete model is required in order to establish whether the theory allows for this sort of phenomena. The goal of this paper is to provide such analysis in a couple of specific physical cases: de Sitter and Schwarzschild spacetimes. We choose the two most-studied examples of deformed relativistic kinematics, and study what happens to particles produced at an interaction vertex that is just outside an observer's horizon. Surprisingly, we find that in both cases, the theory does not allow nonlocality effects to disrupt the notion of horizon. Thus, a strict and ``sharp'' horizon remains present, similar to what is observed in GR.
 
The structure of the paper is as follows. Sec.~\ref{sec:rl_flat} reviews the original proposal of~\cite{AmelinoCamelia:2011bm}, and the geometric construction of~\cite{Relancio:2021ahm}, in the case of a flat spacetime. The generalization to curved spacetimes (carried out in Refs.~\cite{Relancio:2021ahm,Relancio:2020zok,Pfeifer:2021tas}) is described in Sec.~\ref{sec:rl_curved}. In Sec.~\ref{sec:applications} we apply the model to the two simplest examples of spacetimes with horizons, namely Schwarzschild and de Sitter, and study the consequences of RL effects on the notion of horizon. Sec.~\ref{sec:conclusions} contains our conclusions.

\section{Relative Locality in flat spacetime}
\label{sec:rl_flat}

In this section we review how the RL principle was obtained for the first time in~\cite{AmelinoCamelia:2011bm}, and how it  can be derived from a space-time line element with a momentum dependent metric, as done in~\cite{Relancio:2021ahm}. 

\subsection{Original proposal of Relative Locality}
We start by summarizing the original proposal of RL, which begins with the following action (which we write in the particular case of a process involving two incoming and two outgoing particles participating in a 4-valent vertex)~\cite{AmelinoCamelia:2011bm} 
\begin{align}
S^{(2)} \,=&\, \int_{-\infty}^0 d\tau \sum_{i=1,2} \left[x_{-(i)}^\mu(\tau) \dot{p}_\mu^{-(i)}(\tau) + N_{-(i)}(\tau) \left[C(p^{-(i)}(\tau)) - m_{-(i)}^2\right]\right] \nonumber \\   
& + \int^{\infty}_0 d\tau \sum_{j=1,2} \left[x_{+(j)}^\mu(\tau) \dot{p}_\mu^{+(j)}(\tau) + N_{+(j)}(\tau) \left[C(p^{+(j)}(\tau)) - m_{+(j)}^2\right]\right] \nonumber \\
& + \xi^\mu (0)\left[{\cal P}^+_\mu(0) - {\cal P}^-_\mu(0)\right]\,,
\label{S2}
\end{align}
where $\dot{a}\doteq (da/d\tau)$ is a derivative of the variable $a$ with respect to the parameter $\tau$ along the trajectory of the particle, $x_{-(i)}$ ($x_{+(j)}$) are the space-time coordinates of the in-state (out-state) particles, $p^{-(i)}$ ($p^{+(j)}$) their four-momenta, $m_{-(i)}$ ($m_{+(j)}$) their masses, ${\cal P}^-$ (${\cal P}^+$) the total four-momentum of the in-state (out-state) defining the deformed composition law, $C(k)$ the function of momentum-space coordinates $k_\mu$ defining the deformed dispersion relation, $\xi^\mu(0)$ are Lagrange multipliers that implement the energy-momentum conservation in the interaction, and $N_{-(i)}$ ($N_{+(j)}$) are Lagrange multipliers implementing the dispersion relation of in-state (out-state) particles.
 
The variation of the action (\ref{S2}) w.r.t. $x^\mu_{\pm(i)}$ is zero on trajectories that conserve the momenta, $\dot{p}^{\pm(i)}_\mu = 0$. The variation w.r.t. the Lagrange multipliers gives the dispersion relations and the deformed conservation of energy-momentum at the vertex. Finally, the variation w.r.t. the four-momenta $p^{\pm(i)}_\mu$ gives the equations of motion for the coordinates (which produce straight-line trajectories), plus a boundary term which relates the endpoints of these trajectories to the momenta and the Lagrange multipliers $\xi^\mu(0)$:
\be
x_{-(i)}^\mu(0) \,=\, \xi^\nu(0) \frac{\partial {\cal P}^-_\nu}{\partial p^{-(i)}_\mu}(0)\,, \quad\quad\quad
x_{+(j)}^\mu(0) \,=\, \xi^\nu(0) \frac{\partial {\cal P}^+_\nu}{\partial p^{+(j)}_\mu}(0)\,.
\label{eq:rel_loc_original}
\ee

In the undeformed, special-relativistic case, the variables $\xi^\mu(0)$ play the role of the coordinates of the interaction vertex in an inertial reference frame, because $x^\mu_{\pm(i)}(0) = \xi^\mu(0)$. If the conservation law is nonlinearly deformed, then the endpoints of the various particle worldlines will not all coincide with $\xi^\mu(0)$, and will do so in a momentum-dependent fashion. In this case, only an observer placed at the interaction point ($\xi^\mu(0)=0$) will see the interaction as local (all $x^\mu_J(0)$ coincide, being zero). One can choose the Lagrange multiplier $\xi^\mu(0)$ so the interaction will be local only for one class of observers (those whose worldlines cross the interaction event), but all others will see the interaction as nonlocal. Specifically, by extrapolating the trajectories of the particles emitted by the interaction, these observers will reconstruct different emission points for particles with different momenta. The principle of locality, which in a relativistic theory of pointlike particles is implemented by the locality of vertices, is then relaxed. Events, identified by the location of these vertices in space and time, become nonlocal in a way that depends on the state of motion (and location) of the observer. Similarly to what happens to simultaneity in passing from Galilean to SR, locality loses its absoluteness, and the theory is therefore aptly named \emph{Relative Locality.}

\subsection{Relative Locality from geometry}
\label{sec:line_element}

In this Section we will review the proposal first laid out in~\cite{Relancio:2021ahm}, of a classical theory of point particles that implements the RL principle in the context of generalized Hamilton spaces~\cite{miron2001geometry}, starting from a line element in phase space.
 Consider the cotangent bundle $(T^*M)$ of a space-time manifold $M$. Locally, we can define a coordinate system $(x,k)$ on $(T^*M)$, where $x^\mu$ are coordinates on the base manifold  and $k_\mu$ (the momenta) are coordinates on the fiber  (this is called a \textit{local trivialization}). The fiber is a vector space, which at any given point $x$ on the base manifold coincides with the cotangent space $T^*_xM$.
The cotangent bundle to $M$ has a natural structure of symplectic manifold, which can be understood as a special case of Poisson manifold. The Poisson bracket is the natural one induced by the tautological one form of $T^*M$~\cite{miron2001geometry}: one can assume that $x$ and $k$ are canonical coordinates, then  the Poisson brackets are defined by:
\begin{equation}
    \lbrace x^\mu,k_\nu \rbrace\,=\, \delta^\mu_\nu\,. 
\end{equation}

Any fibre bundle $E$, like $T^*M$,  comes equipped with a notion of vertical subspace $V_eE$ at each point $e$ ($V_eE$ is the kernel of the differential of the projection map).  The disjoint union of all vertical spaces is the vertical bundle $VE$. Then, one can define a notion of horizontal bundle as a choice of complementary subbundle to $VE$. This choice is not unique: $E$ does not come equipped with a metric, therefore one cannot define ``perpendicular'' vectors to $V_e$, and any vector subspace of $T_eE$ such that its direct sum with $V_eE$ gives all of $T_eE$ is acceptable. One can associate a connection to this choice, called Ehresmann connection \cite{kobayashi1996foundations}. In~\cite{miron2001geometry} one such connection is introduced for $T^*M$, the \textit{nonlinear connection} $N$. The ``horizontal'' derivative associated to this connection:
\begin{equation}
\frac{\delta}{\delta x^\mu}
=
\frac{\partial}{\partial x^\mu} +  N_{\nu\mu}(x,k)\,\frac{\partial}{\partial k_\mu} \,, 
\label{eq:nonlinear_connection_covariant_derivative}
\end{equation}
moves a point $(x,k) \in T^*M$ in a certain direction, which is not necessarily along a constant-$k_\mu$ curve (all we know is that it is not a constant-$x^\mu$, \textit{i.e.}, a vertical curve.

Within this formalism, we can describe the phase space  of a point-particle in standard GR as follows: the nonlinear connection is chosen to be linear in $k_\mu$:
\begin{equation}
N_{\mu\nu}(x,k)\, = \, k_\rho \Gamma^\rho_{\mu\nu}(x)\,,
\label{eq:nonlinear_connection_in_GR}
\end{equation} 
where now $\Gamma^\rho_{\mu\nu}(x)$ are the Christoffel symbols of a pseudo-Riemannian metric $g_{\mu\nu}(x)$ on $M$. Then certain self-parallel curves generated by the horizontal derivative~(\ref{eq:nonlinear_connection_covariant_derivative}) describe the position and momentum of point particles in GR. 
The projection of these curves on the base manifold are the geodesic~\cite{Relancio:2021ahm} of $g_{\mu\nu}(x)$, and their $k_\mu$ coordinates describe how their momenta (which, unless the metric is homogeneous, is not conserved) evolve. 

The formalism introduced in~\cite{Relancio:2021ahm} allows to generalize the kinematics of point particles to a situation in which the metric  $g_{\mu\nu}$ depends not only on the space-time point $x^\mu$, but also on the momentum (understood as the vertical coordinate $k_\mu$ of $T^*M$). As we will review in the following, this formalism can be generalized to the description of $N$ interacting particles, and, whenever the space-time metric is flat, turns out to be equivalent to the one of RL described in the previous section. This formulation of RL therefore  represents a possible avenue to generalize the original proposal to curved spacetimes, if we allow the metric to depend on both $x^\mu$ and $k_\mu$.

The starting point is the definition of the following line element on $T^*M$ (borrowed from the theory of generalized Hamilton  spaces~\cite{miron2001geometry}):
\begin{equation} 
\mathcal{G}\,=\, g_{\mu\nu}(x,k) dx^\mu dx^\nu+g^{\mu\nu}(x,k) \delta k_\mu \delta k_\nu\,, \qquad \delta k_\mu \,=\, d k_\mu - N_{\nu\mu}(x,k)\,dx^\nu\,, 
\label{eq:line_element_ps} 
\end{equation}
where now we suppose that $g_{\mu\nu}(x,k)$ is a symmetric matrix which can, in principle, depend on both the spacetime coordinates $x^\mu$ and the momenta $k_\mu$, and $g^{\mu\nu}(x,k)$ is its inverse, $g_{\mu\rho}(x,k)g^{\mu\nu}(x,k)=\delta^\nu_\rho$.

A horizontal curve is a geodesic of the metric~(\ref{eq:line_element_ps}) which respects the horizontality condition $\delta k_\mu = 0$, \textit{i.e.:}
\begin{equation}
\frac{d k_\mu}{d \tau} = N_{\nu\mu}(x,k)\, \frac{dx^\nu}{d \tau}\,, 
\label{eq:horizontality_condition}
\end{equation}
where $\tau$ is the proper time (or the affine parameter in case of a massless particle). Under condition~\eqref{eq:horizontality_condition}, the geodesic equation of~(\ref{eq:line_element_ps}) reduces to:
 \begin{equation}
\frac{d^2x^\mu}{d\tau^2}+{H^\mu}_{\nu\sigma}(x,k)\frac{dx^\nu}{d\tau}\frac{dx^\sigma}{d\tau}\,=\,0\,,
\label{eq:horizontal_geodesics_curve_definition}
\end{equation} 
where we will call ${H^\rho}_{\mu\nu}(x,k)$ the ``affine connection of spacetime'', whose explicit expression in terms of $g_{\mu\nu}(x,k)$ and $N_{\mu\nu}(x,k)$ is:
\begin{equation}
\begin{aligned}
{H^\rho}_{\mu\nu}(x,k) &=\,\frac{1}{2}g^{\rho\sigma}(x,k)\left(\frac{\delta g_{\sigma\nu}(x,k)}{\delta x^\mu} +\frac{\delta g_{\sigma\mu}(x,k)}{\delta  x^\nu} -\frac{\delta g_{\mu\nu}(x,k)}{\delta x^\sigma} \right) 
\\
&=\,\frac{1}{2}g^{\rho\sigma}\left(\frac{\partial g_{\sigma\nu}}{\partial x^\mu} +\frac{\partial g_{\sigma\mu}}{\partial  x^\nu} -\frac{\partial g_{\mu\nu} }{\partial x^\sigma} \right) 
\\
&~+
\frac{1}{2}g^{\rho\sigma} \left(N_{\lambda\mu}\frac{\partial g_{\sigma\nu} }{\partial k^\lambda} +
N_{\lambda\nu}\frac{\partial g_{\sigma\mu} }{\partial  k^\lambda} -N_{\lambda\sigma} \frac{\partial g_{\mu\nu} }{\partial k^\sigma} \right)\,.
\end{aligned}
\label{eq:affine_connection_st}
\end{equation} 
It is immediately obvious that, in the standard GR case in which $g_{\mu\nu} =g_{\mu\nu} (x)$ only depends on $x^\mu$, the second term in Eq.~\eqref{eq:affine_connection_st} vanishes, and therefore Eq.~\eqref{eq:horizontal_geodesics_curve_definition} reduces to the standard geodesic equation. Notice also that, in the GR case~\eqref{eq:nonlinear_connection_in_GR}, the horizontality condition~\eqref{eq:horizontality_condition} reduces to 
\begin{equation}
    \frac{d k_\mu}{d \tau} = \, k_\rho \Gamma^\rho_{\mu\nu}(x) \, \frac{dx^\nu}{d \tau}\,,
\label{eq:GR_horizontality}
\end{equation}
which is the equation of motion of momenta that one obtains by varying a point-particle action of the form~\eqref{S2} w.r.t. $k_\mu$, in the standard GR case $C(k) = g^{\mu\nu}(x) k_\mu k_\nu$.  
 
Up to this point, the nonlinear connection $N_{\mu\nu}(x,k) $ and the space-time metric $g_{\mu\nu}(x,k)$ are arbitrary and unrelated to each other. The two can be related if we introduce a Hamiltonian. As shown in~\cite{Relancio:2020rys}, one can introduce a Hamiltonian generating the dynamics of point particles as the function $C(x,k)$, defined as the square of the distance between the point $(x,0)$ and the point $(x,k)$ according to the metric $\mathcal G$. This is the arc length of $\mathcal G$ integrated along a vertical curve and squared:
\begin{equation}
C(x,k) = \left( \int_{(x,0)}^{(x,k)}     \sqrt{g^{\mu\nu}(x,k) dk_\mu dk_\nu} \right)^2 \,. \label{eq:CasimirComoDistancia}
\end{equation}
In GR, where $g^{\mu\nu}(x,k) = g^{\mu\nu}(x)$, the above formula reduces to the quadratic mass Casimir $k_\mu g^{\mu\nu}(x) k_\nu$, while in general it can have a non-quadratic dependence on the momenta.

The Hamiltonian flow generated by the Hamiltonian $\mathcal N \, C(x,k)$ (where $\mathcal N$ is the lapse Lagrange multiplier):
\begin{equation}
    \frac{d x^\mu}{d \tau} = \mathcal N \, \left\{ x^\mu , C(x,k) \right\} = \mathcal N \, \frac{\partial C(x,k)}{\partial k_\mu} \,, \qquad
\frac{d k_\mu}{d \tau} = \mathcal N \, \left\{k_\mu , C(x,k) \right\} = -  \mathcal N \, \frac{\partial C(x,k)}{\partial x^\mu}\,.
\label{eq:HamiltonEqsCasimir}
\end{equation}
Now suppose we assume a horizontality condition with respect to a certain nonlinear connection $N_{\nu\mu}(x,k)$:
\begin{equation}
    \dot k_\mu -  N_{\nu\mu}(x,k)\,\dot x^\nu   = 0 \,.
\end{equation}
The equation above will be, in general, a condition that can be imposed at one point of a solution of Eqs.~(\ref{eq:HamiltonEqsCasimir}), and not on the entire curve (it will be a linear relation between the initial velocities $\dot k_\mu$ and $\dot x^\mu$). For this condition to be preserved along the whole solution, there must be a relation between the nonlinear connection and the Casimir. Specifically, horizontal curves must preserve the Casimir, \textit{i.e.}
\begin{equation}
    \frac{\delta C}{\delta x^\mu} =\frac{\partial C}{\partial x^\mu} +  N_{\mu\nu}(x,k)\,\frac{\partial C}{\partial k_\nu}  = 0 \,,
    \label{eq:ConservationOfCasimir}
\end{equation}
which can be obtained by substituting Eqs.~(\ref{eq:HamiltonEqsCasimir}) into the horizontality condition.
If we now compute the second derivative of the position $x^\mu$ for a horizontal curve, we find:
\begin{equation}
    \frac{d^2 x^\mu}{d\tau^2}+ \frac{\partial N_{\nu\sigma}(x,k)}{\partial k_\mu} \frac{d x^\nu}{d\tau} \frac{d x^\sigma}{d\tau}\,=\,0 \,.
    \label{eq:GeodesicEqFromHamiltonian}
\end{equation}
In principle, Eqs.~\eqref{eq:affine_connection_st}, \eqref{eq:ConservationOfCasimir} and  \eqref{eq:GeodesicEqFromHamiltonian} are sufficient to deduce the form of $N_{\nu\mu}(x,k)$ from that of the metric $g_{\mu\nu}(x,k)$ appearing in the definition of the Casimir~\cite{Relancio:2020rys}
. In practice, this can be done perturbatively, by expanding $g_{\mu\nu}(x,k)$ in powers of the momenta around a classical metric, and imposing~\eqref{eq:affine_connection_st}, \eqref{eq:ConservationOfCasimir} and  \eqref{eq:GeodesicEqFromHamiltonian} order by order~\cite{Pfeifer:2021tas}.

Consider, finally, the case of flat-space RL: the metric depends only on the momenta: $g_{\mu\nu}=g_{\mu\nu}(k)$, and the Hamiltonian~(\ref{eq:HamiltonEqsCasimir}) takes the same form as the Hamiltonian constraint appearing in the original action~(\ref{S2}) for RL. Therefore, the formalism we have described is capable of reproducing the single-particle worldlines of the original model.

However, RL makes little sense as a theory of a single particle: it is when interactions are introduced, that its physical novelty shows. A model aiming to generalize RL to curved spacetimes should at least agree with it in the flat case, when interaction vertices are introduced.

So let us assume, for now, that we are in the spatially-flat case, so that, for a fixed space-time point, the line element~\eqref{eq:line_element_ps} becomes a metric on momentum space:
\begin{equation}
d\sigma^2\,=\,g^{\mu\nu}(k) d k_\mu d k_\nu\,.
\label{eq:line_element_ps_flat2} 
\end{equation}
We are interested in a model that respects the \textit{special} relativity principle, \textit{i.e.}, the invariance of physics under change of inertial observers. As discussed in detail in~\cite{Carmona:2019fwf}, the way to implement this in our formalism is to suppose that the metric~(\ref{eq:line_element_ps_flat2}) is a maximally symmetric (\textit{i.e.}, Minkowski or (Anti-)de Sitter, with a ``momentum-space cosmological constant'' $\Lambda$ with the dimensions of an energy), \textit{as a metric on momentum space}. Then, there is a Lorentz subgroup of isometries, which reflects onto (Lorentzian) rotations of the worldline coordinates, implementing thereby rotations and Lorentz boosts of the observers. Moreover, one has a four-dimensional subgroup of translation isometries. These are translations \textit{in momentum space,} so their physical interpretation is less obvious. To understand it, notice that a translation subgroup acting transitively on momentum space can be represented as a binary composition of momentum coordinates $\oplus : (p,q) \to p \oplus q$. The condition that $\oplus$ is an isometry implies that:
\begin{equation}
g^{\mu\nu}(q) d q_\mu d q_\nu\,=\, g^{\mu\nu}\left(p\oplus q\right)d \left(p\oplus q\right)_\mu d \left(p\oplus q\right)_\nu\,,
\label{eq:line_element_isometry} 
\end{equation}
which implies, for the transformation rule of the inverse metric $g_{\mu\nu}(k)$, that:
\begin{equation}
g_{\mu\nu}\left(p\oplus q\right) \,=\,\frac{\partial \left(p\oplus q\right)_\mu}{\partial q_\rho} g_{\rho\sigma}(q)\frac{\partial \left(p\oplus q\right)_\nu}{\partial q_\sigma}\,.
\label{eq:composition_isometry}
\end{equation} 
The isometries of the momentum metric do not have to be isometries of the full phase-space metric. In general:
\begin{equation}
g_{\mu\nu}\left(p\oplus q\right) dx^\mu dx^\nu + g^{\mu\nu}\left(p\oplus k\right) d\left(p\oplus k\right)_\mu d \left(p\oplus k\right)_\nu \neq g_{\mu\nu}(p) dx^\mu dx^\nu + g^{\mu\nu} dk_\mu d k_\nu\,,
\end{equation} 
because $g_{\mu\nu}\left(p\oplus q\right) dx^\mu dx^\nu $ is not necessarily equal to $g_{\mu\nu}(p) dx^\mu dx^\nu $. However, one can always build an isometry of the full metric by asking that the coordinates transform at the same time as the momenta:
\begin{equation}
g_{\mu\nu}\left(p\oplus q\right) d\xi^\mu d\xi^\nu + g^{\mu\nu}\left(p\oplus q\right) d\left(p\oplus q\right)_\mu d \left(p\oplus k\right)_\nu = g_{\mu\nu}(p) dx^\mu dx^\nu + g^{\mu\nu}(p) d p_\mu d p_\nu\,,
\end{equation} 
where now
\begin{equation}\label{CoordinateOplusTransformationLaw}
\frac{\partial x^\mu}{\partial \xi^\nu} =  \frac{\partial (p \oplus k)_\nu}{\partial k_\mu} \qquad \Rightarrow \qquad x^\mu = \frac{\partial (p \oplus k)_\nu}{\partial k_\mu} \xi^\nu + \text{\it const.} \,,
\end{equation}
and the integration constant above can be set to zero without loss of generality. This relation coincides with Eq.~(\ref{eq:rel_loc_original}), the expression of the relationship between the coordinates of the interaction vertex $\xi^\mu(0)$ and the endpoint of each particle's worldline $x^\mu(0)$ in standard (flat-space) RL. This observation inspired one of us~\cite{Relancio:2021ahm} to find a formulation of interaction vertices in RL that works in the case of non-flat spacetimes too.

Notice first that the operation $\oplus$ is covariant under boost transformations, and can be used as the basic building block in order to construct interaction vertices. In fact, $\oplus$ reduces to the standard momentum sum ``$+$'' when the momentum metric is flat, and therefore, in the curved (A)dS case, it represents a generalization of the composition law of momenta. For a 4-valent vertex with two particles in the initial and two particle in the final state, the conservation of momentum can be written as:
\begin{equation}\label{2to2momentumConservation}
    (p(0) \oplus q(0) )_\mu = (k(0)  \oplus r(0))_\mu  \,,
\end{equation}
where we conventionally parametrized the particle worldlines so that their parameter time is zero at the interaction.

What we need is a principle/geometrical law that reproduces, in the flat limit, Eq.~(\ref{eq:rel_loc_original}) for the endpoint coordinates, and Eq.~(\ref{2to2momentumConservation}) for the momentum conservation law. In~\cite{Relancio:2021ahm}, the following principle was proposed: in a scattering process in which no new particles are created (as long as one focuses on the classical theory, only such processes are allowed), each particle changes its momentum and its endpoint coordinates, in passing from the initial to the final state, according to a momentum-space isometry of the type $q \to q \oplus p $ or $q \to p \oplus q $ (together with transformations of the type~(\ref{CoordinateOplusTransformationLaw}) for the endpoint coordinates.\footnote{We would like to underline that, as an intermediate step in the derivation of the rules of the model, in Ref.~\cite{Relancio:2021ahm} the discussion initially assumed that each particle in a classical vertex retains its identity before and after the interaction, as happens for example in an elastic collision of macroscopic objects. This was essential in order to identify each particle in the final state with one and only one particle in the initial state, so that their momenta could be related by an isometry. This assumption, of course, cannot be retained when Quantum Mechanics is taken into account, because identical particles will obey the Bose--Einstein or Fermi--Dirac statistics, and there will be no way to identify the particles in the initial and final state. When Relativistic Quantum Mechanics (\textit{i.e.} QFT) effects are taken into account, not even the number of particles is conserved between the initial and final state, so there can be no such identification, in principle. At any rate, the assumption of Ref.~\cite{Relancio:2021ahm} was just an intermediate step in identifying the final, consistent law that the initial and final particle momenta satisfy. Once this law has been derived, as we will see, it will not depend on any identification between initial and final particles, and will be generalizable to (Relativistic) Quantum Mechanical models.}
If that is the case, then the phase-space line element, calculated at the coordinates and momenta of one particle in the initial state (\textit{e.g.} $q_\mu$ and $z^\mu$), will be the same as that calculated at the coordinates and momenta of the same particle in the final state (\textit{e.g.} $r_\mu$ and $w^\mu$):
\begin{equation}\label{ConservationOfLineElement}
g_{\mu\nu}\left(q\right) dz^\mu dz^\nu + g^{\mu\nu}\left(q\right) dq_\mu d q_\nu  = g_{\mu\nu}\left(r\right) dw^\mu dw^\nu + g^{\mu\nu}\left(r\right) dr_\mu d r_\nu \,.
\end{equation}
Such an equation guarantees that, for this particle, Eq.~(\ref{CoordinateOplusTransformationLaw}) is valid. However, one incurs into a problem when trying to impose the same condition on the other particle (with initial and final momenta, respectively, $p_\mu$ and $k_\mu$): if the composition law $\oplus$ is not symmetric (which is the case in most applications), then the relation~(\ref{CoordinateOplusTransformationLaw}) will not be valid for this particle.

The way out of this problem is laid out in Ref.~\cite{Relancio:2021ahm}: one has 
 to enlarge our configuration space to that of $N$ particles, and introduce a line element on this $8N$-dimensional space. This line element involves all the particles in the initial and final state of a process. For example, for two particles:
\begin{equation}
\mathcal{G}_2\,=\, G_{AB}(P) \, dX^A dX^B + G^{AB}(P) d P_A d P_B\,,
\label{eq:line_element_ps2}
\end{equation}
where $ G_{AB}(P) $ is an 8-dimensional metric
\begin{equation}
G_{AB}(P)\,=\,
\begin{pmatrix}
g^{LL}_{\mu\nu}(p,q) & g^{LR}_{\mu\nu}(p,q) \\
g^{RL}_{\mu\nu}(p,q) & g^{RR}_{\mu\nu}(p,q) 
\end{pmatrix}\,,
\label{eq:8-metric}
\end{equation}  
$X^A=(y^\mu,z^\mu)$, $P_A=(p_\mu,q_\mu)$, and $A$, $B$ run from $0$ to $7$. Again, $G^{AB}(P)$ is the inverse of $G_{AB}(P)$. In the SR limit, when $\Lambda$ goes to infinity, one finds
\begin{equation}
G_{AB}^{(SR)}\,=\,
\begin{pmatrix}
\eta_{\mu\nu} &0 \\
0& \eta_{\mu\nu}
\end{pmatrix}\,.
\label{eq:8-metric_SR}
\end{equation}  
This means that the line element in spacetime,
\begin{equation}
ds^2_2\,=\, G_{AB}(P) \, dX^A dX^B\,,
\label{eq:line_element_sp2}
\end{equation}
should be 0, 1, or 2, depending if we are considering two massless particles, one massless and other massive, or two massives ones (the worldlines of massive particles are parametrized with respect to proper time).
In order to have a symmetric metric, the components of (\ref{eq:line_element_ps2}) have to satisfy: 
\begin{equation}
g^{LL}_{\mu\nu}(p,q)\,=\,g^{LL}_{\nu\mu}(p,q)\,,\qquad g^{LR}_{\mu\nu}(p,q)\,=\,g^{RL}_{\nu\mu}(p,q)\,,\qquad g^{RR}_{\mu\nu}(p,q)\,=\,g^{RR}_{\nu\mu}(p,q)\,.
\end{equation}

Now we are ready to introduce the conservation law that will control the physics of the interaction vertex: the 8-dimensional line element should be the same before and after the interaction:
\begin{equation}
 G_{AB}(P) \, dX^A dX^B + G^{AB}(P) d P_A d P_B
 = 
  G_{AB}(K) \, dV^A dV^B + G^{AB}(K) d K_A d K_B\,,
\label{eq:line_element_ps2_conservation}
\end{equation}
where $X^A$ and $P_A$ are the coordinates and momenta of the incoming particles,
while $V^A$ and $K_A$ are those  of the outgoing particles.

Since we are considering the geodesic trajectories of particles, following horizontal curves (for which $\delta k_\mu=0$), the line element $\mathcal{G}_2$ reduces to 
\begin{equation}
ds^2_2\,=\, G_{AB}(P) dX^A dX^B\,.
\label{LineElementOfParticleG2}
\end{equation}
Therefore, since the composition law is an isometry of the 8-dimensional momentum metric, and the same metric is used for both space-time and momentum line elements by definition, then the composition law must be also an isometry of the space-time line element.

To ensure that the line element is conserved according to~(\ref{eq:line_element_ps2_conservation}) and the outgoing coordinates/momenta are related to the incoming ones by an isometry, we can introduce an intermediate state satisfying 
	\begin{equation}
 G_{AB}(P) dX^A dX^B\,=\,2 g_{\mu\nu}\left(p\oplus q\right) d\xi^\mu d\xi^\nu\,.
 \label{eq:line_element_vertex2}
\end{equation}
The factor 2 appears since we are asking that the two particles have the same vertex of the interaction. Otherwise, we do not obtain the SR limit for which interactions are local~\cite{Relancio:2021ahm}. 

The calculations are simplified by introducing an 8-dimensional tetrad to depict the metric~\eqref{eq:8-metric}
\begin{equation}\Phi^A_B(p,q)\,=\,
\begin{pmatrix}
\varphi^{(L)\alpha}_{(L)\mu}(p,q) & \varphi^{(L)\alpha}_{(R)\mu} (p,q) \\
\varphi^{(R)\alpha}_{(L)\mu}(p,q) & \varphi^{(R)\alpha}_{(R)\mu}  (p,q)
\end{pmatrix}\,,
\label{eq:8-tetrad}
\end{equation}
such that 
	\begin{equation}
 G_{AB}(P) \,=\, \Phi^C_A(p,q)
\eta_{CD} \Phi^D_B(p,q)  \,,
\end{equation}
where 
	\begin{equation}
\eta_{CD} \,=\, \begin{pmatrix}
\eta_{\alpha\beta}&0\\
0& \eta_{\alpha\beta}
\end{pmatrix}  \,.
\end{equation}
The tetrad splits as follows in terms of the block-components of the metric~\eqref{eq:8-metric}
\begin{equation}
\begin{split}
g^{LL}_{\mu\nu}(p,q)\,&=\,\varphi^{(L)\alpha}_{(L)\mu}(p,q)\eta_{\alpha\beta}\varphi^{(L)\beta}_{(L)\nu}(p,q)+\varphi^{(R)\alpha}_{(L)\mu}(p,q)\eta_{\alpha\beta}\varphi^{(R)\beta}_{(L)\nu}(p,q)\,,\\
g^{LR}_{\mu\nu}(p,q)\,&=\,g^{RL}_{\nu\mu}(p,q)\,=\,\varphi^{(L)\alpha}_{(L)\mu}(p,q)\eta_{\alpha\beta}\varphi^{(L)\beta}_{(R)\nu}(p,q)+\varphi^{(R)\alpha}_{(L)\mu}(p,q)\eta_{\alpha\beta}\varphi^{(R)\beta}_{(R)\nu}(p,q)\,,\\
g^{RR}_{\mu\nu}(p,q)\,&=\,\varphi^{(L)\alpha}_{(R)\mu}(p,q)\eta_{\alpha\beta}\varphi^{(L)\beta}_{(R)\nu}(p,q)+\varphi^{(R)\alpha}_{(R)\mu}(p,q)\eta_{\alpha\beta}\varphi^{(R)\beta}_{(R)\nu}(p,q)\,.
\end{split}
\label{eq:metric_tetrad}
\end{equation}

Since we want to recover the metric for a single particle when there is only one momentum, we impose 
\begin{equation}
\varphi^{(L)\alpha}_{(L)\nu}(p,0)\,=\,\varphi^{\alpha}_{\mu}\left(p\right)\,,\qquad   \varphi^{(L)\alpha}_{(R)\nu} (0,q)\,=\, \varphi^{(R)\alpha}_{(L)\nu}(p,0)\,=\,0\,,\qquad   \varphi^{(R)\alpha}_{(R)\nu} (0,q)\,=\,\varphi^{\alpha}_{\mu}\left(q\right) \,,
\label{eq:varphi_L2}
\end{equation}
where $\varphi^{\alpha}_{\mu}\left(p\right)$ is the (inverse of the) tetrad in momentum space 
\begin{equation}
 g_{\mu\nu}(k)\,=\,\varphi^{\alpha}_{\mu}(k)\eta_{\alpha\beta}\varphi^{\beta}_{\nu}(k)\,.
\end{equation}
Then the composition law will be an isometry of the 8-dimensional metric if
\begin{equation}
\begin{split}
\varphi^{\alpha}_{\mu}\left(p\oplus q\right)\,&=\,\frac{\partial \left(p\oplus q\right)_\mu}{\partial p_\nu}\varphi^{(L)\alpha}_{(L)\nu}(p,q) +\frac{\partial \left(p\oplus q\right)_\mu}{\partial q_\nu} \varphi^{(L)\alpha}_{(R)\nu} (p,q) \,,\\
\varphi^{\alpha}_{\mu}\left(p\oplus q\right)\,&=\,\frac{\partial \left(p\oplus q\right)_\mu}{\partial p_\nu}\varphi^{(R)\alpha}_{(L)\nu}(p,q) +\frac{\partial \left(p\oplus q\right)_\mu}{\partial q_\nu} \varphi^{(R)\alpha}_{(R)\nu} (p,q) \,,
\end{split}
\label{eq:varphi_L}
\end{equation}
is satisfied.

Then, the desired RL condition~\eqref{eq:rel_loc_original} can be obtained from  Eq.~\eqref{eq:line_element_vertex2} 
\begin{equation}
\varphi^{\alpha}_{\nu}\left(p\oplus q\right)\,=\,\frac{\partial y^\mu }{\partial \xi^\nu}\varphi^{(L)\alpha}_{(L)\mu}(p,q) +\frac{\partial z^\mu }{\partial \xi^\nu} \varphi^{(L)\alpha}_{(R)\mu} (p,q)\,=\,\frac{\partial y^\mu }{\partial \xi^\nu}\varphi^{(R)\alpha}_{(L)\mu}(p,q) +\frac{\partial z^\mu }{\partial \xi^\nu} \varphi^{(R)\alpha}_{(R)\mu} (p,q) \,.
\label{eq:tetrad_tetrads}
\end{equation}
Therefore, due to the conditions~\eqref{eq:varphi_L} one can see that the previous equation is satisfied if
\begin{equation}
\frac{\partial y^\mu }{\partial \xi^\nu}\,=\,\frac{\partial \left(p\oplus q\right)_\nu}{\partial p_\mu}\,,\qquad\frac{\partial z^\mu }{\partial \xi^\nu}\,=\,\frac{\partial \left(p\oplus q\right)_\nu}{\partial q_\mu} \,,
\label{eq:rl_flat_coord}
\end{equation}
which is consistent with~\eqref{eq:rel_loc_original}.

As discussed in~\cite{Relancio:2021ahm}, this prescription can be applied to a generic deformed relativistic kinematics (in particular, for $\kappa$-Poincaré in the basis considered in~\cite{Carmona:2019vsh} and the most generic kinematics at first order in the deformation parameter considered in~\cite{Carmona:2012un}) and extended to any number of particles. This allows us to construct a metric in phase space for a multi-particle system, and also define a space-time noncommutativity through the (inverse of the) tetrads~\eqref{eq:8-tetrad}. 

\section{Relative Locality in curved spacetime}
\label{sec:rl_curved}

In the previous section we have summarized how the principle of RL can be obtained from a line element in phase space with a momentum dependent metric. In~\cite{Relancio:2021ahm} it was claimed that this procedure can be generalized to the case of a curved spacetime, i.e., a metric depending on both space-time and momentum coordinates. In this section we extend the aforementioned framework so to include a generic space-time curvature. 

In~\cite{Relancio:2020zok,Pfeifer:2021tas} the results of~\cite{Carmona:2019fwf} were extended in order to consider a deformed kinematics and a curved spacetime within the same framework. To that aim, it is mandatory to consider the cotangent bundle geometry discussed above (Eq.~\eqref{eq:line_element_ps}). The cotangent-bundle metric tensor $g_{\mu\nu}(x,k)$, depending on space-time coordinates, can be constructed in terms of the spacetime tetrad field and the metric of  momentum space, $\bar{g}$, explicitly~\cite{Relancio:2020zok, Pfeifer:2021tas}
\begin{equation}
g_{\mu\nu}(x,k)\,=\,e^\alpha_\mu(x) \bar{g}_{\alpha\beta}(\bar{k})e^\beta_\nu(x)\,,
\label{eq:definition_metric_cotangent}
\end{equation}
where $\bar{k}_\alpha=e^\nu_\alpha (x) k_\nu$, and $e^\alpha_\mu(x)$ denotes the tetrad of spacetime. As discussed in~\cite{Relancio:2020zok}, when constructing the metric in this way we assure that the momentum space (fiber in the cotangent bundle) is maximally symmetric if the starting metric $\bar{g}$ also is. Moreover, in~\cite{Pfeifer:2021tas}  this construction  was rederived by lifting the symmetries for flat spacetimes to the curved case. 

As shown in \cite{Pfeifer:2021tas}, the most general form of the metric, in which the construction of a deformed kinematics in a curved space-time background is allowed, is  a momentum space metric whose Lorentz isometries are linear transformations in the momenta, i.e., a metric of the form
\begin{equation}
\bar g_{\alpha\beta}(k)\,=\,\eta_{\alpha\beta} f_1 (k^2)+\frac{k_\alpha k_\beta}{\Lambda^2} f_2(k^2)\,,
\label{eq:lorentz_metric}
\end{equation}
where $\Lambda$ is the high-energy scale parametrizing the momentum deformation of the metric and kinematics (not to be confused with the cosmological constant, which in the present paper is not used, preferring to refer to the horizon radius $l$ instead),  and $\eta=\mathrm{diag}(1,-1,-1,-1)$. From Eq.~\eqref{eq:definition_metric_cotangent} one obtains the following metric in the cotangent bundle when a curvature in spacetime is present
\begin{equation}
g_{\mu \nu}(x,k)\,=\,a_{\mu \nu}(x)f_1 (\bar{k}^2)+\frac{k_\mu  k_\nu}{\Lambda^2} f_2(\bar{k}^2)\,,
\label{eq:lorentz_metric_curved}
\end{equation}
where $a_{\mu \nu}(x) = e^\alpha_\mu(x) \eta_{\alpha\beta} e^\beta_\nu(x)$ is the curved space-time metric.

For this metric, it is possible to define a composition law as an isometry~\cite{Relancio:2020zok,Pfeifer:2021tas}.  That is, for a fixed space-time point, the line element~\eqref{eq:line_element_ps} becomes 
\begin{equation}
d\sigma^2\,=\,g^{\mu\nu}(x,k) d k_\mu d k_\nu\,=\,\bar g^{\alpha\beta}(\bar k) d \bar k_\alpha d \bar k_\beta\,.
\label{eq:line_element_ps_curved2} 
\end{equation}
Therefore, there is a simple generalization of~\eqref{eq:composition_isometry} 
\begin{equation}
\bar g_{\alpha\beta}\left(\bar p\oplus \bar q\right) \,=\,\frac{\partial \left(\bar p\oplus\bar q\right)_\alpha}{\partial \bar q_\gamma} \bar g_{\gamma\delta}(\bar q)\frac{\partial \left(\bar p\oplus  \bar q\right)_\beta}{\partial \bar q_\delta}\,.
\label{eq:composition_isometry_curved}
\end{equation} 

Then, we consider the generalization of the line element~\eqref{eq:line_element_ps2} for the curved spacetime case   
\begin{equation}
\mathcal{G}_2\,=\, G_{\mu\nu}(X,\bar{P}) dX^\mu dX^\nu+G^{\mu\nu}(X,\bar{P}) \delta P_\mu \delta P_\nu\,,
\label{eq:line_element_ps2_curved}
\end{equation}
where $\bar{P}_\mu=(\bar{p}_\mu,\bar{q}_\mu)=(\bar{e}_\mu^\nu(y) p_\nu ,\bar{e}_\mu^\nu(z) q_\nu)$, $\delta P_\mu = \left(\delta p_\mu , \delta q_\mu ) = (dp_\mu-N_{\nu\mu}(y,p)dy^\nu , dq_\mu-N_{\nu\mu}(z,q)dz^\nu\right)$, and    $X^A=(y^\mu,z^\mu)$.

This generalization allows us to define the RL of interaction in a curved spacetime. As before, particles follow horizontal curves, for which the last term of Eq.~\eqref{eq:line_element_ps2_curved} vanishes. Then, taking into account that the composition law is an isometry of the 8-dimensional metric, we can define an intermediate state, which is the generalization for a curved spacetime of Eq.~\eqref{eq:line_element_vertex2}
	\begin{equation}
 ds^2_2\,=\,G_{\mu\nu}(X(0),\bar{P}(0)) dX^\mu(0) dX^\nu(0)\,=\,2 g_{\mu\nu}\left(\xi,\bar p\oplus \bar q\right) d\xi^\mu d\xi^\nu \,.
\label{eq:line_element_vertex2_curved}
\end{equation}

In order to obtain the RL principle for curved spacetimes,  we can define a tetrad as in Eq.~\eqref{eq:8-tetrad} for a metric which depends on positions and momenta
\begin{equation}\Phi^A_B(y,z,p,q)\,=\,
\begin{pmatrix}
e^\beta_\mu(y)\varphi^{(L)\alpha}_{(L)\beta}(\bar p,\bar q) & e^\beta_\mu(z)\varphi^{(L)\alpha}_{(R)\beta} (\bar p,\bar q) \\
e^\beta_\mu(y)\varphi^{(R)\alpha}_{(L)\beta}(\bar p,\bar q) & e^\beta_\mu(z) \varphi^{(R)\alpha}_{(R)\beta}  (\bar p,\bar q)
\end{pmatrix}\,.
\label{eq:8-tetrad_curved}
\end{equation}
Now, we can write~\eqref{eq:line_element_vertex2_curved} as a function of the tetrads, viz.
\begin{equation}
e^\rho_\nu (\xi)\varphi^{\alpha}_{\rho}\left(\bar p\oplus \bar q\right)\,=\, \frac{\partial y^\mu }{\partial \xi^\nu} e^\sigma_\mu (y) \varphi^{(L)\alpha}_{(L)\sigma}(\bar p,\bar q) +\frac{\partial z^\mu }{\partial \xi^\nu} e^\sigma_\mu (z)  \varphi^{(L)\alpha}_{(R)\sigma} (\bar p,\bar q)\,=\,  \frac{\partial y^\mu }{\partial \xi^\nu} e^\sigma_\mu (y) \varphi^{(R)\alpha}_{(L)\sigma}(\bar p,\bar q) +\frac{\partial z^\mu }{\partial \xi^\nu} e^\sigma_\mu (z)  \varphi^{(R)\alpha}_{(R)\sigma} (\bar p,\bar q) \,.
\label{eq:rel_loc_curved_eqs}
\end{equation}

Since we have made the replacement $k\to \bar k$, we can simply rewrite Eq.~\eqref{eq:tetrad_tetrads} (when taking into account~\eqref{eq:rl_flat_coord}) as a function of the barred variables
\begin{equation}
\varphi^{\alpha}_{\nu}\left(\bar p\oplus \bar q\right)\,=\,\frac{\partial \left(\bar p\oplus \bar q\right)_\nu}{\partial \bar p_\mu} \varphi^{(L)\alpha}_{(L)\mu}(\bar p,\bar q) +\frac{\partial \left(\bar p\oplus \bar q\right)_\nu}{\partial \bar q_\mu} \varphi^{(L)\alpha}_{(R)\mu} (\bar p,\bar q)\,=\,\frac{\partial \left(\bar p\oplus \bar q\right)_\nu}{\partial \bar p_\mu} \varphi^{(R)\alpha}_{(L)\mu}(\bar p,\bar q) +\frac{\partial \left(\bar p\oplus \bar q\right)_\nu}{\partial \bar q_\mu} \varphi^{(R)\alpha}_{(R)\mu} (\bar p,\bar q) \,.
\label{eq:tetrad_tetrads_curved}
\end{equation}
Again, since the composition law is an isometry of the metric, we ask the line element in spacetime to be invariant under such transformation. By substituting these expressions in~\eqref{eq:rel_loc_curved_eqs} we can obtain a simple solution, that does not depend on the choice of the tetrad of the 8-dimensional metric (as in the flat case), describing the RL principle for curved spacetimes
\begin{equation}
\frac{\partial y^\mu }{\partial \xi^\nu} e^\sigma_\mu (y) \,=\,e^\rho_\nu (\xi) \frac{\partial \left(\bar p\oplus \bar q\right)_\rho}{\partial \bar p_\sigma}\,,\qquad \frac{\partial z^\mu }{\partial \xi^\nu} e^\sigma_\mu (z) \,=\,e^\rho_\nu (\xi) \frac{\partial \left(\bar p\oplus \bar q\right)_\rho}{\partial \bar q_\sigma} \,.
\label{eq:rl_flat_curved}
\end{equation}
This system of equations can be solved order by order in the deformation parameter $\Lambda$ describing the deformed kinematics. In order to see the first-order correction, we can make the ansatz 
\begin{equation}
 y^\mu \,\approx\ \xi^\nu \left(\delta^\mu_\nu +A^\mu_\nu (y,z,p,q)  \right) \,,\qquad  z^\mu \,\approx\ \xi^\nu \left(\delta^\mu_\nu +B^\mu_\nu (y,z,p,q)  \right) \,,
\label{eq:solution_1}
\end{equation}
where $A^\mu_\nu (y,z,p,q)$ and $B^\mu_\nu (y,z,p,q)$ are corrections that vanish when $\Lambda \to \infty$.
Also, we can expand the composition law in a similar way, and in particular its derivatives, in the following way
\begin{equation}
  \frac{\partial\left(\bar p\oplus \bar q\right)_\nu}{\partial \bar p_\mu} \,\approx\  \delta^\mu_\nu +C^\mu_\nu (y,z,p,q) \,,\qquad    \frac{\partial\left(\bar p\oplus \bar q\right)_\nu}{\partial \bar q_\mu} \,\approx\  \delta^\mu_\nu +D^\mu_\nu (y,z,p,q)  \,,
\label{eq:comp_1}
\end{equation}
where $C^\mu_\nu (y,z,p,q)$ and $D^\mu_\nu (y,z,p,q)$ vanish too as $\Lambda \to \infty$.
Using Eqs.~\eqref{eq:solution_1}-\eqref{eq:comp_1} into Eq.~\eqref{eq:rl_flat_curved} we find for the left particle
\begin{equation}
  \left(\delta^\mu_\nu +A^\mu_\nu \right)  e^\sigma_\mu (y) \,=\, \left(e^\lambda_\nu (y)+\frac{\partial e^\lambda_\nu (y)}{\partial y^\tau}\left(\xi^\tau-y^\tau\right)\right)\left(\delta^\sigma_\lambda +C^\sigma_\lambda\right)\,,
\end{equation}
so 
\begin{equation}
 A^\mu_\nu e^\sigma_\mu (y) \,=\,-\frac{\partial e^\sigma_\nu (y)}{\partial y^\tau} y^\rho A^\tau_\rho  + e^\lambda_\nu (y)  C^\sigma_\lambda \,.
 \label{eq:solution_y}
\end{equation}
Analogously, one easily finds for the right particle 
\begin{equation}
 B^\mu_\nu e^\sigma_\mu (z) \,=\,-\frac{\partial e^\sigma_\nu (z)}{\partial z^\tau} z^\rho B^\tau_\rho  + e^\lambda_\nu (z)  D^\sigma_\lambda \,.
  \label{eq:solution_z}
\end{equation}
Although in the previous expressions  $y$ and $z$ appear, both can be replaced by $\xi$ since we are keeping only the lowest-order contribution in $\Lambda$ to the RL effect. By solving the previous equations one then finds the coordinates of the particles involved in a scattering before the interaction (the extension to the particles after the collision is straightforward). Since it is an algebraic equation, it is in principle solvable for any tetrad of spacetime (we will see an example in the following section). Moreover, it is important to note that, since everything is defined through a metric in the cotangent bundle, all the previous results are invariant under space-time diffeomorphisms.   

As mentioned in the introduction,  an action with some nonlocal variables (defined by the space-time tetrad) was considered in~\cite{Cianfrani:2014fia}. Here we obtained a different description of RL for curved spacetimes that relies on the canonically conjugated variables $(x,k)$. This is quite natural in this scheme, since our starting point is precisely a cotangent bundle geometry defined and constructed with these variables.

Let us stop for a moment to consider the structure of Eqs.~(\ref{eq:rl_flat_curved}) and their perturbative versions, Eqs.~(\ref{eq:solution_y}) and ~(\ref{eq:solution_z}). They can be understood as relations between the initial coordinates $y^\mu$, $z^\mu$ of each particle's worldline, and the ``interaction coordinates'' $\xi^\mu$. The structure of the equations is the same as in the vanilla, flat-space RL theory: the initial coordinates of all particles tend to coincide with the interaction coordinates as $\xi^\mu$ tends to zero, \textit{i.e.}, near the origin of a coordinate system. The nonlocality of the interaction vertex increases linearly with $\xi^\mu$, and can become observably large at cosmological distances. Once an observer (\textit{i.e.}, a coordinate system) is chosen, each particle worldline will have their initial coordinates $y^\mu$ or $z^\mu$ as initial conditions, and will thereafter propagate according to the horizontality condition. In a perturbative approach, this means that they will propagate along geodesics of the background metric  (that is, $\lim_{\Lambda \to \infty} e^\sigma_\mu$), with corrections of order at least $\Lambda^{-1}$. One can easily provide an example of a model in which the particle dispersion relation does not depend on $\Lambda$ (it is undeformed), and then the whole RL effect is concentrated in the spreading of the initial coordinates of each particles at the interaction vertex.
This observation leads naturally to a question: what happens to causal relations in this theory?
 
One of the main distinctive features of GR, that distinguishes it from SR, is the presence of horizons and causally disconnected patches of spacetime. While in Minkowski space an inertial observer can receive a signal from anywhere in spacetime, as long as they wait a sufficiently long time, the same is not true, for example, in de Sitter spacetime: there is a causal diamonds of spacetime points that can send signals to an inertial observer, and points outside it can never do that, no matter what the observer does. This is exemplified by the fact that, in an expanding universe with a sufficiently large positive cosmological constant, distant galaxies will eventually cross the cosmological horizon and be cut off from communication with Earth, so that we can only observe their light up to a certain point in their evolution.
Instead, in a theory in which high-energy processes have a growing degree of nonlocality which increases with the distance from the observer, one can expect a particle that comes from an interaction vertex outside of the causal patch of the observer to appear to them as if it was emitted inside of their cosmological horizon.
Similarly, in a Schwarzschild spacetime, one could imagine that a vertex taking place just inside of the event horizon (in the sense that the interaction coordinates place it inside of the horizon) could produce some particles that appear outside of the horizon and are therefore able to reach an outside observer.
The particular structure of the RL theory (\textit{i.e.}, the form of the composition law and the geometry of momentum space) might mitigate (or prevent altogether) this sort of phenomenon, giving rise to a weak notion of horizon (or keeping the GR notion unmodified); however, there is nothing in the current assumptions of the theory that guarantees that. In the remainder of the paper, we will investigate whether the GR notion of horizon is preserved, modified, or broken, in two spacetimes which have a notion of horizon: de Sitter and Schwarzschild.

\section{Particular examples}
\label{sec:applications}

In this section we apply the proposed model for RL in curved spacetimes to the particular cases of a de Sitter universe and an eternal (Schwarzschild) black hole. We will focus on the existence of the cosmological and event horizon, respectively.  As test cases, we will consider two of the most studied deformed kinematics in DSR: $\kappa$-Poincaré~\cite{Borowiec2010} and Snyder~\cite{Battisti:2010sr} models.

\subsection{de Sitter scenario}

\begin{figure}[h!]
    \centering
    \includegraphics[width=0.9\textwidth]{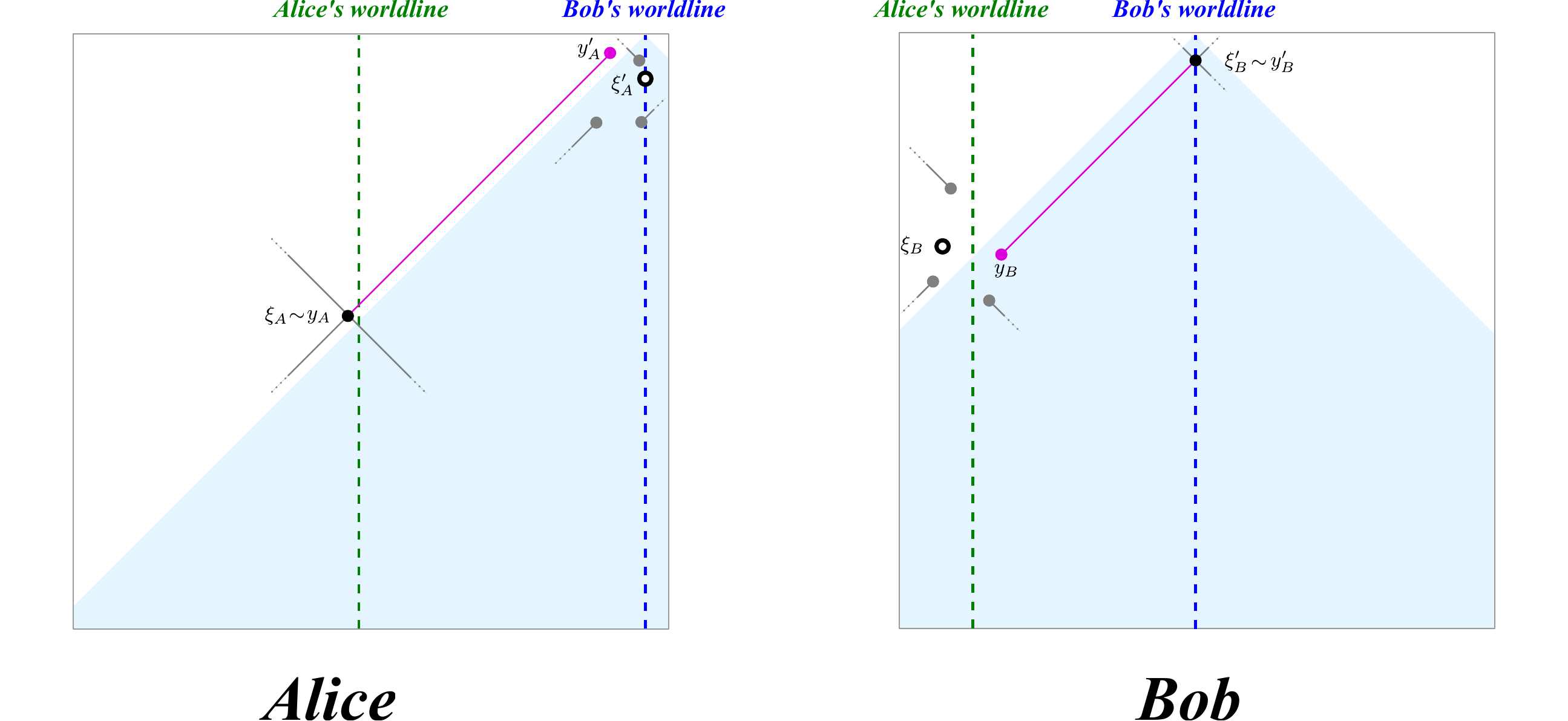}
    \caption{Penrose--Carter diagrams of de Sitter spacetime, in two coordinate system centred on the comoving observers Alice (left) and, respectively, Bob (right). 
    If the RL effects break the GR notion of cosmological horizon, the two observers end up disagreeing on whether the purple particle can reach observer Bob.     
    According to Alice (left), the purple particle is produced outside of the causal patch of Bob, and can therefore never reach him. From Bob's perspective (right), the same particle is produced inside of his causal patch, and will therefore eventually reach him. For simplicity, we assumed that Alice and Bob are comoving observers, and Alice sits in the south pole of her coordinate system, while Bob sits in the north pole of his own coordinate system. }
    \label{dSfigure}
\end{figure}

\begin{figure}[h!]
    \centering
    \includegraphics[width=0.9\textwidth]{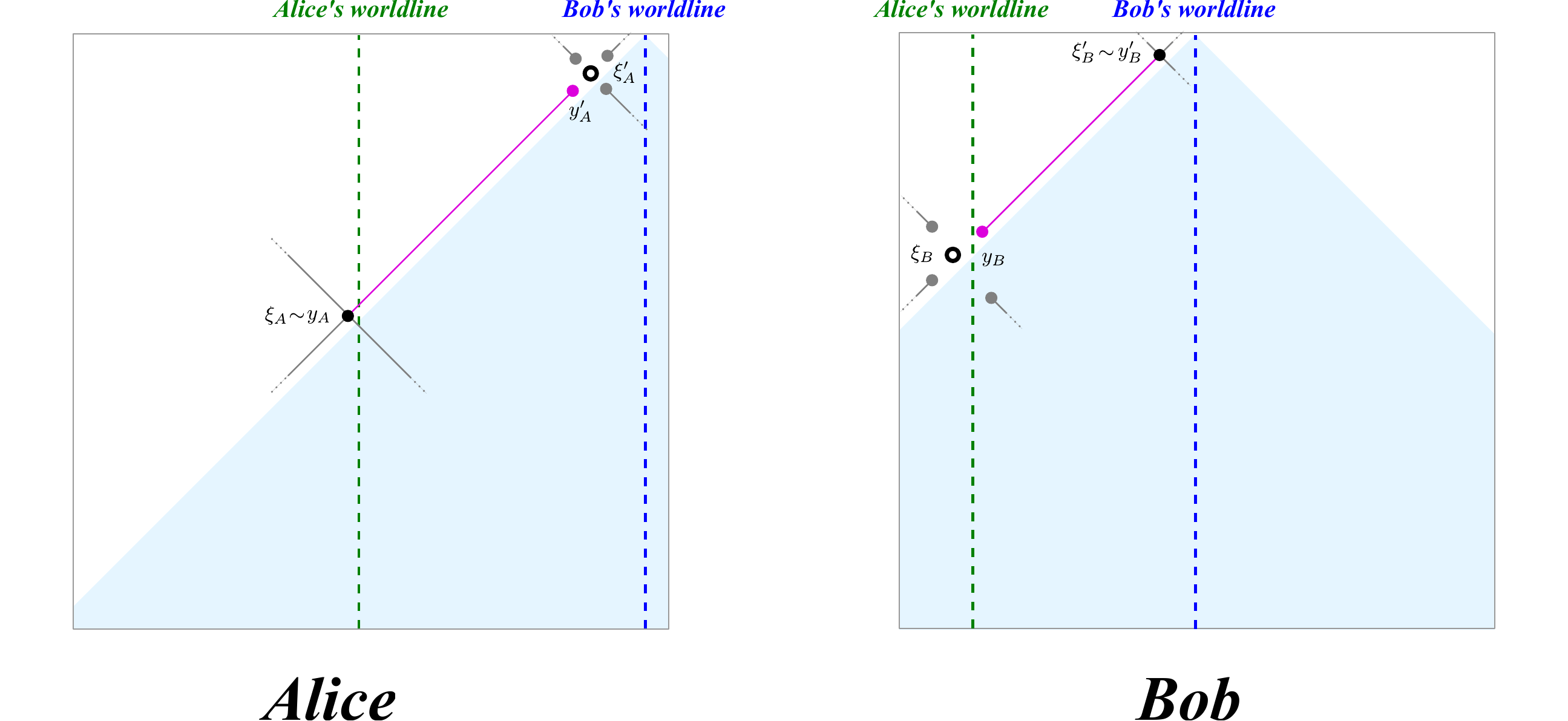}
    \caption{In another scenario, the starting point of the purple particle in Bob's reference frame, $y_B$, despite being shifted from the location of the interaction coordinates $\xi_B$, does not lie inside of Bob's causal cone, and is therefore unable to reach Bob. Both observers agree on whether the purple particle reaches Bob.}
    \label{dSfigure2}
\end{figure}

For the de Sitter case, we are interested in collisions near the cosmological horizon of an observer, in order to check whether the RL    principle allows some particles to ``break out'' of the horizon. Let us assume that a collision takes place outside the horizon of observer Bob, but very close to it (Fig.~\ref{dSfigure} and~\ref{dSfigure2}), so the interaction coordinates $\xi_B$ in Bob's coordinate system are outside of his causal patch.
Observer Alice, whose worldline gets her very close to the interaction vertex, sees all of the worldlines of the  particles as originating from the same point $\xi_A$, outside of the horizon. Since Bob is far from the collision coordinates, the starting points of some of the particles, \textit{i.e.} $y_B$, might actually appear to be inside of his horizon, and he will be able to detect them (Fig.~\ref{dSfigure}). The detection can be simply modeled as a second interaction vertex whose interaction coordinates $\xi'_B$ are close to Bob's worldline. Then, in his reference frame, the detection point $y_B$ of the particle sent by Alice coincides with $\xi'_B$. From Alice's perspective, the interaction coordinates $\xi'_A$ of the detection vertex is inside of Bob's causal patch, but the detection coordinate $y'_A$ of the particle she sent is outside of it (as was the whole worldline of the particle).  This is one possibility; however, there are other scenarios in which this effect is confined to a small region, and then the cosmological horizon would become ``fuzzy'', or even disappear entirely, as illustrated in Fig.~\ref{dSfigure2}. In this scenario, although the interaction vertices, as seen from distant observers, are still nonlocal, the emission/detection points of the particles never cross over the cosmological horizon, which is therefore still valid as a concept. As we will show in the continuation, in the two  models of deformed kinematics that we have considered, the latter scenario is realized (at least within a de Sitter or Schwarzschild spacetime), and the cosmological horizon of de Sitter, and the event horizon of Scwharzschild, still make sense as separations between causally disconnected regions.

Since we want to focus on a region close to the horizon, we need to use some coordinates that are regular at it. In particular, we will use the Kruskal  coordinates~\cite{Gibbons:1977mu}
\begin{equation}
a_{UV}\,=\,-\frac{2 l^2}{(1- U V)^2}\,,\qquad a_{\theta \theta }\,=\,\frac{l^2(1+ U V)^2}{(1- U V)^2}\,,\qquad a_{\phi \phi }\,=\, \sin^2 \theta \frac{l^2(1+ U V)^2}{(1- U V)^2}\,,
\label{eq:deSitter_metric}
\end{equation}
where $l$ is the horizon radius. A particular choice of tetrad is
\begin{equation}
e^U_U\,=\,e^U_V\,=\,e^V_U\,=\,-e^V_V\,=\,\frac{l}{1-UV}\,, \qquad e^\theta_\theta \,=\, l \frac{1+UV}{1-UV}\,,\qquad e^\phi_\phi\,=\,l \frac{1+UV}{1-UV} \sin{\theta}\,.
\label{eq:kruskal_tetrad}
\end{equation}
In these coordinates, the horizon of a comoving observer sitting at the south (north) pole corresponds to the line $U=0$ ($V=0$). Moreover, both $U$ and $V$ vary within the range $[-1,1]$. Indeed, these Kruskal coordinates can be easily related to the well-known flat slicing ones~\cite{Weinberg:1972kfs} 
 \begin{equation}
ds^2\,=\, -dt^2+ e^{2 t/l}dr^2\,,
\label{eq:flat_slicing}
\end{equation}
through
 \begin{equation}
\frac{r}{l}\,=\, \frac{1+ U V}{1- U V}\,,\qquad e^{2t/l}\,=\,- \frac{ U }{V}\,.
\label{eq:kruskal_fs}
\end{equation}

For the sake of simplicity, we will use the following assumptions. 
Since we are only interested in interactions taking place near the horizon ($U=0$ or $V=0$), we make a choice of the vertex coordinates such that $\theta=\phi=0$.  Also, we will study photons moving only along the $U$ $(V)$ direction, so $p_\theta=q_\theta=p_\phi=q_\phi=0$, and, therefore, $p_V=0$ $(p_U=0)$, and   $q_U=0$ $(q_V=0)$. This choice is motivated by the fact that, for simplicity, we will choose two models of deformed relativistic kinematics in which the dispersion relation in undeformed~\cite{Borowiec2010,Battisti:2010sr}, \textit{i.e.}, it takes the usual expression of GR
\begin{equation}
    C(k)\,=\,\bar k_\mu \eta^{\mu\nu}\bar k_\nu\,=\,-k_U k_V \frac{\left(1- UV\right)^2}{l^2}\,,
\end{equation}
for $k_\theta=k_\phi=0$ (notice that the momenta $k_U$ and $k_V$, which are conjugate to the dimensionless coordinates $U$ and $V$,  are dimensionless themselves, so the energy-squared dimensions of the equation above are given to it entirely by the $\ell^{-2}$ factor). For massless particles, $ C(k)=0$, so $k_U=0$ or  $k_V=0$.  All the nontriviality of the chosen models~\cite{Borowiec2010,Battisti:2010sr} is contained in the deformed momentum composition laws.

We are interested in collisions near the cosmological horizon, so the radial collision coordinate $\xi_r$ will be close to $l$. We can define a small dimensionless parameter $\epsilon$ as:
\begin{equation}
\xi_r = l(1+\epsilon) \,.
\end{equation}
$\epsilon$ measures the difference between $\xi_r$ and $l$ in units of $l$,
so one of  the corresponding Kruskal interaction coordinates, $\xi_U$  or $\xi_V$, will be small when $\epsilon$ is small. This allows us to deduce the following relationship between $\xi_U$ and $\xi_V$ from the first expression of~\eqref{eq:kruskal_fs}
\begin{equation}\label{RelationXiUXiV}
    \xi_U \, \xi_V\, \simeq \,\frac{\epsilon}{2}\,.
\end{equation}
When considering  an observer sitting at the north pole, whose horizon is at $V=0$, the interaction point will be close to the horizon if $\xi_V = \epsilon/2 \, \xi_U$, where $\xi_U$ is arbitrary.

\subsubsection{\texorpdfstring{$\kappa$}{k}-Poincaré kinematics}
The composition law of $\kappa$-Poincaré in the classical basis reads~\cite{Borowiec2010}
\begin{equation}
    (p\oplus q)_0\,=\, p_0\, \Pi(q)+\Pi^{-1}(p)\left(q_0+\frac{\vec{p}\cdot \vec{q}}{\Lambda}\right)\,,\qquad    (p\oplus q)_i\,=\, p_i \,\Pi(q)+q_i\,,
    \label{eq:dcl_bc}
\end{equation}
with
\begin{equation}
   \Pi(k)\,=\, \frac{k_0}{\Lambda}+\sqrt{1+\frac{k_0^2-\vec{k}^2}{\Lambda^2}}\,,\qquad  \Pi^{-1}(k)\,=\, \left(\sqrt{1+\frac{k_0^2-\vec{k}^2}{\Lambda^2}}-\frac{k_0}{\Lambda}\right)\left(1-\frac{\vec{k}^2}{\Lambda^2}\right)^{-1}\,.
\end{equation}
At first order in $\Lambda$ one finds
\be
(p\oplus q)_0 \,= \,p_0 +q_0 + \frac{\vec{p}\cdot \vec{q}}{\Lambda} \,,\qquad (p\oplus q)_i \,= \,p_i \left(1+\frac{q_0}{\Lambda}\right) +q_i \,.
\label{DCLkappa-1}
\ee
It can be checked that this is the most general composition law  at first order in $\Lambda$, preserving isotropy, associativity, and compatible with an undeformed dispersion relation and the condition~\eqref{eq:composition_isometry_curved}.

Substituting this composition law with the space-time tetrad~\eqref{eq:kruskal_tetrad} into Eqs.~\eqref{eq:solution_y}-\eqref{eq:solution_z}, and taking into account~\eqref{RelationXiUXiV}, one finds for the two particles
\be
y^U\,= \, \xi_U \left(1 + \frac{ q_U (1+\xi_U^2)}{2 \Lambda l}\right)-\frac{\epsilon}{4 l \xi_U \Lambda}\left((1+\xi_U^2)(q_V+2q_U \xi_U^2)\right)\,,\qquad y^V\,= \, \frac{\epsilon}{2  \xi_U}\left(1+\frac{q_V+2q_U \xi_U^2}{2 l  \Lambda} \right)-\frac{ q_U  \xi_U}{2 \Lambda l}\,,
\label{eq:dsrel_y1}
\ee
\be
z^U\,= \, \xi_U \left(1 + \frac{ p_U -p_V}{4 \Lambda l}(2- \epsilon)\right)\,,\qquad z^V\,= \,\frac{\epsilon}{2  \xi_U} \left(1-\frac{ (p_U -p_V)}{2 \Lambda l}\right) \,.
\label{eq:dsrel_z1}
\ee
We only displayed the $(U,V)$ components since they are the ones we are interested in. 
The emission coordinates $y$ and $z$ of the particles differ  from the interaction coordinates $\xi$ only by a correction of order  (dimensionless momenta)$/(\Lambda \ell)$, because $	|\xi_U| < 1 $ by definition. The whole perturbative scheme is based on the assumption that the (dimensional) momenta are all much smaller than $\Lambda$, which, in terms of the dimensionless momenta $p_U$, $p_V$ etc. means that (dimensionless momenta)$/(\Lambda \ell) \ll 1$.

Now consider the case in which $p_V = q_U =0$, \textit{i.e.}, the photon with starting point $y^\mu$ (and momentum $p_\mu$) propagates in the direction parallel to the horizon (the $U$ direction), while  the photon that starts in $z^\mu$ (with momentum $q_\mu$) propagates in the direction perpendicular to it (the $V$ direction). Then 
Eq.~\eqref{eq:dsrel_y1}  reduces to 
\be
y^U\,=  \xi_U  -\frac{\epsilon}{ l\Lambda} \left( \frac{1+\xi_U^2}{4 \, \xi_U} \right) \, q_V \,,\qquad y^V\,= \, \frac{\epsilon}{2  \xi_U} \left( 1 + \frac{q_V}{2 \, l  \, \Lambda} \right)\,,
\label{eq:dslrel_y1_limit}
\ee
in which we see that the coordinate $y^V$, which is the one perpendicular to the horizon, gets corrected by a term that can rescale it by a little bit, but not change its sign, unless  $q_V /(2\ell \Lambda) < -1$, which would bring us out of the perturbative scheme and would invalidate the whole perturbative analysis.

We can focus instead on the right particle, which is emitted at the coordinates $z$. An observer sitting at the south pole will have their horizon at $U =0$, so we need to exchange the role of the $U$ and $V$ components in the relation $\xi_U=\epsilon/2\, \xi_V$, so that Eqs.~(\ref{eq:dsrel_y1})and~(\ref{eq:dsrel_z1}) now read:
\be
y^U\,= \, \frac{\epsilon}{2  \xi_V}\left(1+\frac{q_U+2q_V \xi_V^2}{2 l  \Lambda} \right)-\frac{ q_V  \xi_V}{2 \Lambda l}\,,\qquad y^V\,= \,\xi_V \left(1 + \frac{ q_V (1+\xi_V^2)}{2 \Lambda l}\right)-\frac{\epsilon}{4 l \xi_V \Lambda}\left((1+\xi_V^2)(q_U+2q_V \xi_V^2)\right) \,,
\label{eq:dsrel_y1uv}
\ee
\be
z^U\,= \, \frac{\epsilon}{2  \xi_V} \left(1-\frac{ (p_V -p_U)}{2 \Lambda l}\right)\,,\qquad z^V\,= \, \xi_V \left(1 + \frac{ p_V -p_U}{4 \Lambda l}(2- \epsilon)\right)\,.
\label{eq:dsrel_z1uv}
\ee
In order to make the right particle parallel to the horizon we have to impose, like before, $p_V = q_U =0$.
Just like $y^V$ before,  the coordinate $z^U$ cannot change sign due to the correction term because $ \left|\frac{p_U}{2 \Lambda l}\right|  \ll 1$, and the GR horizon is again respected.

\subsubsection{Snyder kinematics}

We will now consider a different model, based on  the composition law of Snyder in the Maggiore representation~\cite{Battisti:2010sr}
\be
(p\oplus q)_\mu \,= \,p_\mu \left(\sqrt{1+ \frac{ q_\mu q_\nu \eta^{\mu\nu}}{\Lambda^2}}+\frac{p_\mu q_\nu \eta^{\mu\nu}}{\Lambda^2 \left(1+\sqrt{1+ p_\mu p_\nu \eta^{\mu\nu}/\Lambda^2}\right)}  \right) + q_\mu \,,
\label{DCLSnyder-1}
\ee
which at second order reads
\be
(p\oplus q)_\mu \,= \,p_\mu \left(1+ \frac{ q_\mu q_\nu \eta^{\mu\nu}}{2\Lambda^2}+\frac{p_\mu q_\nu \eta^{\mu\nu}}{2\Lambda^2}  \right) + q_\mu \,.
\label{DCLSnyder}
\ee
Note that this is not the most general composition law at second order compatible with linear Lorentz covariance and satisfying~\eqref{eq:composition_isometry_curved}, with a generic de Sitter momentum metric at second order. However, the result is qualitatively independent of the composition used so, for the sake of simplicity, we will use along this discussion the aforementioned composition law.

Using the composition law~\eqref{DCLSnyder} with the space-time tetrad~\eqref{eq:kruskal_tetrad} into Eqs.~\eqref{eq:solution_y}-\eqref{eq:solution_z}, and taking into account~\eqref{RelationXiUXiV}, one finds
\begin{align}
y^U\,=& \, \xi_U \left(1 + \frac{1}{4 \Lambda^2 l^2}\left( p_U q_U \xi_U^2- p_V q_U-2(p_U+q_U)q_V\right)\right)\nonumber\\
&+\frac{\epsilon}{8 \Lambda ^2 l^2 \xi_U }  \left(\xi_U ^2 (7 p_U q_V+5 p_V q_U+8 q_U q_V)-3 p_U q_U \xi_U^4-p_V q_V\right)\,, \nonumber\\
y^V\,=& \, \frac{\epsilon}{2  \xi_U}\left( 1+\frac{1  }{4 \Lambda ^2 l^2 }\left(3 p_U q_U \xi_U^2-p_U q_V-2 q_U (p_V+q_V)\right) \right)-\frac{\xi_U p_U q_U}{4 \Lambda^2l^2}\,,
\label{eq:dsrel_y2}
\end{align}
\begin{align}
z^U\,=& \, \xi_U \left(1 + \frac{p_U\left( (p_U+2 q_U) \xi_U^2- p_V-2 q_V\right)}{4 \Lambda^2 l^2}\right) \nonumber \\
&+\frac{\epsilon}{8 \Lambda ^2 l^2 \xi_U }\left(p_V-3 p_U \xi_U\right) \left(\left(\xi_U (p_U+2 q_U)\right)-p_V-2 q_V\right)\,,  \nonumber \\
z^V\,= &\, \frac{\epsilon}{2  \xi_U}\left(1+\frac{(p_U+2 q_U) \left(3 p_U \xi_U^2-p_V\right)}{4 \Lambda ^2 l^2 }\right)-\frac{\xi_U p_U (p_U+ 2q_U)}{4 \Lambda^2l^2}\,,
\label{eq:dsrel_z2}
\end{align}
where we eliminated coordinate $\xi_V$ by replacing it with $ \epsilon/2 \, \xi_U$. 
Again, we only displayed the $(U,V)$ components since they are the ones we are interested on in the following.  Similar expressions, changing the role of $U$ and $V$, can be obtained when considering the horizon of an observer sitting at the south pole, in which case $\xi_U=\epsilon/2\, \xi_V$.

Now consider the case in which $p_V = q_U =0$, \textit{i.e.}, the photon with starting point $y^\mu$ (and momentum $p_\mu$) propagates in the direction parallel to the horizon (the $U$ direction), while  the photon that is emitted at $z^\mu$ (with momentum $q_\mu$) propagates in the direction perpendicular to it (the $V$ direction). Then 
Eq.~\eqref{eq:dsrel_y2} reduces to:
\begin{equation}
y^U = \, \xi_U \left[ 1 - \frac{p_U \, q_V}{2 \Lambda^2 l^2} \left(1 +	\frac{7}{4} \epsilon \, \xi_U   \right) \right]  \,, \qquad
y^V = \, \frac{\epsilon}{2  \xi_U}\left( 1 -  \frac{ p_U \, q_V  }{4 \Lambda ^2 l^2 } \right) \,,
\label{eq:dsrel_y2_limit}
\end{equation}
The sign of the coordinate $y^V$, which discriminates between particles that are emitted inside ($y^V <0$) and outside ($y^V > 0$) of the horizon, cannot be changed by the correction term  because $\left| \frac{p_U \, q_V  }{4 \Lambda ^2 l^2 } \right| \ll 1$. Therefore, if the interaction vertex is inside the horizon ($\epsilon$<0), the particle that goes in the direction of Bob can reach him, because $y^U <0$, while, if the vertex is outside, the particle can never reach Bob.

Now, changing the role of $U$ and $V$, we can consider the right particle instead of the left one, assuming the interaction point is near the horizon of an observer sitting at the south pole, in which case $\xi_U=\epsilon/2\, \xi_V$. Eqs.~(\ref{eq:dsrel_y2}) and~(\ref{eq:dsrel_z2}) turn into:
\begin{align}
y^U\,=& \,\frac{\epsilon}{2  \xi_V}\left( 1+\frac{1  }{4 \Lambda ^2 l^2 }\left(3 p_V q_V \xi_V^2-p_V q_U-2 q_V (p_U+q_U)\right) \right)-\frac{\xi_V p_V q_V}{4 \Lambda^2l^2} \,, \nonumber \\
y^V\,=& \, \xi_V \left(1 + \frac{1}{4 \Lambda^2 l^2}\left( p_V q_V \xi_V^2- p_U q_V-2(p_V+q_V)q_U\right)\right)\nonumber \\
&+\frac{\epsilon}{8 \Lambda ^2 l^2 \xi_V }  \left(\xi_V ^2 (7 p_V q_U+5 p_U q_V+8 q_V q_U)-3 p_V q_V \xi_V^4-p_U q_U\right)\,,
\label{eq:dsrel_y2uv}
\end{align}
\begin{align}
z^U\,=& \,\frac{\epsilon}{2  \xi_V}\left(1+\frac{(p_V+2 q_V) \left(3 p_V \xi_V^2-p_U\right)}{4 \Lambda ^2 l^2 }\right)-\frac{\xi_V p_V (p_V+ 2q_V)}{4 \Lambda^2l^2} \,,  \nonumber \\
z^V\,= &\, \xi_U \left(1 + \frac{p_V\left( (p_V+2 q_V) \xi_V^2- p_U-2 q_U\right)}{4 \Lambda^2 l^2}\right)+\frac{\epsilon}{8 \Lambda ^2 l^2 \xi_V }\left(p_U-3 p_V \xi_V\right) \left(\left(\xi_V (p_V+2 q_V)\right)-p_U-2 q_U\right)\,,
\label{eq:dsrel_z2uv}
\end{align}
now imposing the  condition $p_V=q_U=0$, the equation for $z^U$ reduces to:
\begin{align}
z^U = \,\frac{\epsilon}{2  \xi_V}\left(1-\frac{ q_V p_U} {2 \Lambda ^2 l^2 }\right)  \,,  \qquad
z^V =\, \xi_U  +\frac{\epsilon}{8 \Lambda ^2 l^2 \xi_V }  \left(2 \xi_V q_V p_U  - p_U^2\right)\,.
\label{eq:dsrel_z2uv_limit}
\end{align}
 Like before, the coordinate $z^U$ cannot change sign because of the correction term, which  is such that $\left| \frac{ q_V p_U} {2 \Lambda ^2 l^2 } \right| \ll 1$.

\subsection{Schwarzschild scenario}

Also in the case of a Schwarzschild black hole, the Kruskal coordinates~\cite{Mukhanov:2007zz} are the best choice to discuss a region around the horizon. In these coordinates the metric is:
\begin{equation}
a_{UV}\,=\,-\frac{32 M^3}{r}e^{-r/2M}\,,\qquad a_{\theta \theta }\,=\,r\,,\qquad a_{\phi \phi }\,=\, r \sin^2 \theta \,,
\label{eq:Schw_metric}
\end{equation}
where $M$ is the mass of the black hole, and $r$ is defined implicitly by the equation
\begin{equation}
U V\,=\, \left(1-\frac{r}{2 M} \right) e^{r/2M}\,.
\label{eq:rUV}
\end{equation}
A particular choice of tetrad is
\begin{equation}
e^U_U\,=\,e^U_V\,=\,e^V_U\,=\,-e^V_V\,=\,\frac{4 M^{3/2}}{r^{1/2}} e^{-r/4 M}\,, \qquad e^\theta_\theta \,=\, r\,,\qquad e^\phi_\phi\,=\,r \sin{\theta}\,.
\label{eq:kruskal_tetrad_Schw}
\end{equation}
In these coordinates, the usual event horizon corresponds to (half of) the line $V=0$, while the black hole exterior, which is the only accessible region to an observer outside of the black hole, is the quadrant $U>0$, $V<0$. The curvature singularity is placed at $UV=1$. 

For the sake of simplicity, we will use the following assumptions. 
Since we are only interested in interactions taking place near the horizon ($U=0$ or $V=0$), we make the choice of the vertex coordinates such that $\theta=\phi=0$.  Like before, we will study photons moving only along the $U(V)$ direction, so $p_\theta=q_\theta=p_\phi=q_\phi=0$, and, therefore, $p_V=0\,(p_U=0)$, and    $q_V=0\,(q_U=0)$. Again, this is a consequence of the fact that the dispersion relation of the models ~\cite{Borowiec2010,Battisti:2010sr} is undeformed:
\begin{equation}
    C(k)\,=\,\bar k_\mu \eta^{\mu\nu}\bar k_\nu\,=\,-k_U k_V  \frac{r e^{r/2 M}}{16 M^3}\,,
\end{equation}
for $k_\theta=k_\phi=0$. For massless particles, $ C(k)=0$, so $k_U=0$ or  $k_V=0$. 

\begin{figure}[h!]
    \centering
    \includegraphics[width=0.9\textwidth]{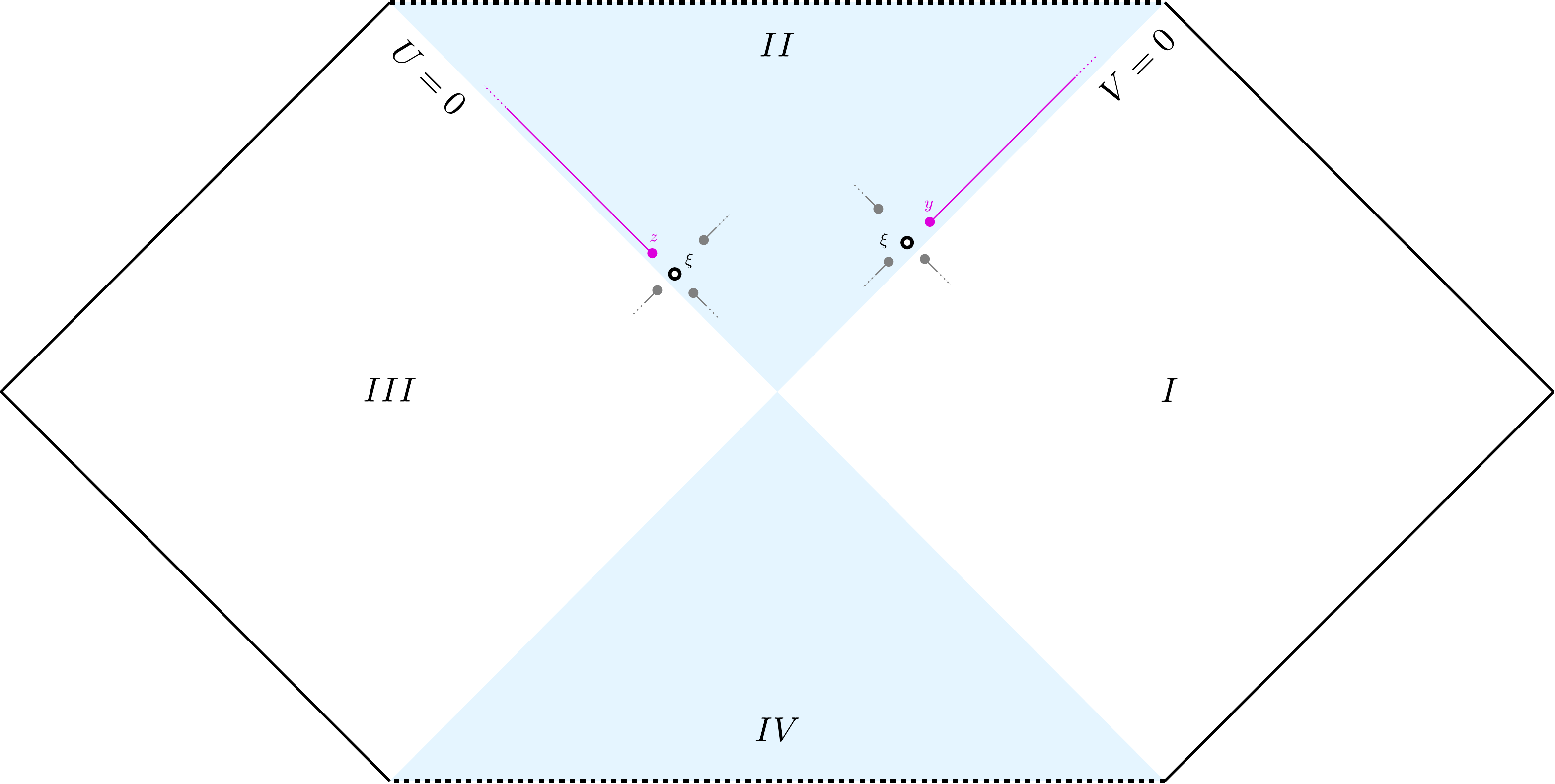}
    \caption{The Schwarzschild scenario is showed in this illustrative Penrose--Carter diagram, valid for  both the $\kappa$-Poincar\'e and the Snyder kinematics. The $I$ and $III$ quadrant are equally good  descriptions of an (idealized, eternal) black hole exterior. If the interaction coordinate $\xi$  is close to the black hole horizon of quadrant $I$, then the emission coordinate of particle $y$ (the one that propagates parallel to the horizon, \textit{i.e.} in the $U$ direction), is in the black hole interior, quadrant $II$. If the interaction coordinate $\xi$  is close to the black hole horizon of quadrant $III$, then the emission coordinate of particle $z$ (the one that propagates in the direction $V$), is in the black hole interior, quadrant $II$. }
    \label{FigSchwarzschild}
\end{figure}

Moreover, we are interested in collisions near the event horizon,
\begin{equation}
    r=2M(1+\epsilon) \,, \label{eq:NearEventHorizon}
\end{equation}
so either $U$ or $V$ is small (notice that the Kruskal extension of Schwarzschild spacetime has four quadrants, see Fig.~\ref{FigSchwarzschild}, and both the first and third quadrant are good descriptions of a black hole exterior. The black hole event horizon is at $U=0$ in the first quadrant, and at $V=0$ in the third quadrant). 
Eq.~(\ref{eq:NearEventHorizon})  allows us to deduce the following relationship between $U$ and $V$ from the first expression of~\eqref{eq:rUV}
\begin{equation}
   V\,=\,-\frac{e \epsilon}{U}\,.
   \label{RelationXiUXiV2}
\end{equation}

\subsubsection{\texorpdfstring{$\kappa$}{k}-Poincaré kinematics}

Substituting this composition law with the space-time tetrad~\eqref{eq:kruskal_tetrad_Schw} into Eqs.~\eqref{eq:solution_y}-\eqref{eq:solution_z}, and taking into account~\eqref{RelationXiUXiV2}, one finds 
\begin{align}
y^U\,= &\, \xi_U \left(1 + \frac{ q_U (e+\xi_U^2)}{4 \sqrt{2 e} \Lambda M}\right) + \frac{\epsilon  \left(q_U \xi_U^4+4 e q_U \xi_U^2+2 e q_V \xi_U^2+2 e^2 q_V\right)}{8 \sqrt{2 e} \Lambda  M \xi_U}\,,\nonumber\\ 
y^V\,=& \, -\frac{e \epsilon}{  \xi_U}\left(1+\frac{2 q_U \xi_U^2+e q_V}{4 \sqrt{2 e} \Lambda  M}\right)-\frac{ \sqrt{e}  q_U \xi_U}{ 4 \sqrt{2} \Lambda M }\,,
\label{eq:screl_y1}
\end{align}
\begin{align}
z^U\,=& \, \xi_U \left(1+  \frac{ \sqrt{e}  (p_U -p_V)}{4 \sqrt{2} \Lambda M}(1+\epsilon)\right) \,,\nonumber\\ z^V\,=& \, -\frac{e \epsilon}{  \xi_U}\left(1+\sqrt{\frac{e}{2}}\frac{ (p_V-p_U)}{4 \Lambda  M}\right)\,.
\label{eq:screl_z1}
\end{align}

If now consider the case in which $p_V = q_U =0$, this is what the coordinate $y$ reduces to:
\begin{align}
y^U =  \xi_U   + \sqrt{\frac e 2 } \frac{\epsilon \, q_V }{4  \Lambda  M \xi_U} \left( \xi_U^2 +  e \right) \,, \qquad  
y^V = -\frac{e \epsilon}{  \xi_U}\left(1+\sqrt{\frac e 2 }  \frac{ q_V}{4  \Lambda  M}\right) \,.
\label{eq:screl_y1_limit}
\end{align}
$y^V$ is the relevant coordinate, here. Its sign determines which side of the horizon the particle is emitted from, and the correction term, $\left| \sqrt{\frac e 2 }  \frac{ q_V}{4  \Lambda  M} \right| \ll 1 $, cannot hope to change this sign. The Schwarzschild horizon of GR is preserved.

Exchanging roles between left and right particle and $U$ and $V$ coordinates, one finds:
\begin{align}
y^U\,= &\,-\frac{e \epsilon}{  \xi_V}\left(1+\frac{2 q_V \xi_V^2+e q_U}{4 \sqrt{2 e} \Lambda  M}\right)-\frac{ \sqrt{e}  q_V \xi_V}{ 4 \sqrt{2} \Lambda M }\,,\nonumber\\ 
y^V\,=& \,  \xi_V \left(1 + \frac{ q_V (e+\xi_V^2)}{4 \sqrt{2 e} \Lambda M}\right) + \frac{\epsilon  \left(q_V \xi_V^4+4 e q_V \xi_V^2+2 e q_U \xi_V^2+2 e^2 q_U\right)}{8 \sqrt{2 e} \Lambda  M \xi_V}\,,
\label{eq:screl_y1UV}
\end{align}
\begin{align}
z^U\,=& \,-\frac{e \epsilon}{  \xi_V}\left(1+\sqrt{\frac{e}{2}}\frac{ (p_U-p_V)}{4 \Lambda  M}\right)  \,,\nonumber\\ z^V\,=& \, \xi_V \left(1+  \frac{ \sqrt{e}  (p_V -p_U)}{4 \sqrt{2} \Lambda M}(1+\epsilon)\right)\,,
\label{eq:screl_z1UV}
\end{align}
and, imposing $p_V = q_U =0$, the $z$ coordinates tend to:
\begin{equation}
z^U = \,-\frac{e \epsilon}{  \xi_V}\left(1+\sqrt{\frac{e}{2}}\frac{p_U}{4 \Lambda  M}\right)  \,,
\qquad
z^V = \, \xi_V \left(1 - \frac{ \sqrt{e} \, p_U}{4 \sqrt{2} \Lambda M}(1+\epsilon)\right)\,.
\label{eq:screl_z1UV_limit}
\end{equation}
$z^U$ cannot change sign because of the correction term $\left| \sqrt{\frac{e}{2}}\frac{p_U}{4 \Lambda  M} \right| \ll 1$, so the GR horizon is preserved in this case too.

\subsubsection{Snyder kinematics}

With these assumptions, and using the composition law~\eqref{DCLSnyder} with the space-time tetrad~\eqref{eq:kruskal_tetrad_Schw} into Eqs.~\eqref{eq:solution_y}-\eqref{eq:solution_y}, one finds
\begin{align}
y^U\,= &\,\xi_U\left(1 + \frac{p_U q_U \xi_U^2-e p_V q_U -2 e (p_U+q_U) q_V}{32 M^2 \Lambda^2}\right)+\frac{\epsilon  \left(-2 e \xi_U^2 (7 p_U q_V+5 p_V q_U+8 q_U q_V)+3 p_U q_U \xi_U^4+2 e^2 p_V q_V\right)}{64 \Lambda ^2 M^2 \xi_U}\,,\nonumber\\
y^V\,= &\, -\frac{e \epsilon}{\xi_U}\left(1+\frac{3 p_U q_U \xi_U^2-e (p_U q_V+2 q_U (p_V+q_V))}{32 \Lambda ^2 M^2}\right)-\frac{e \xi_U p_U q_U}{32 M^2 \Lambda^2 }\,.
\label{eq:rel_ys}
\end{align}
\begin{align}
z^U\,= &\,\xi_U\left(1 + \frac{p_U \left((p_U+2q_U)\xi^2_U-e(p_V+2q_V)\right)}{32 M^2 \Lambda^2}\right)\nonumber
\\
&+\frac{\epsilon  \left(-4 e \xi_U^2 (2 p_U p_V+3 p_U q_V+p_V q_U)+3 p_U \xi_U^4 (p_U+2 q_U)+2 e^2 p_V (p_V+2 q_V)\right)}{64 \Lambda ^2 M^2 \xi_U}\,,\nonumber
\\ z^V\,= &\, -\frac{e \epsilon}{\xi_U}\left(1+\frac{(p_U+2 q_U) \left(3 p_U \xi_U^2-e p_V\right)}{32 \Lambda ^2 M^2}\right)-\frac{e \xi_U p_U(p_U+2 q_U)}{32 M^2 \Lambda^2 }\,.
\label{eq:rel_zs}
\end{align}
If now consider the case in which $p_V = q_U =0$, the $y$ 
coordinates become
\begin{equation}
y^U =  \xi_U\left(1 - \frac{ e p_U  q_V}{32 M^2 \Lambda^2} (2 + 7 \epsilon) \right)\,,\qquad y^V = -\frac{e \epsilon}{\xi_U}\left(1 - \frac{ e p_U q_V}{32 \Lambda ^2 M^2}\right) \,,
\label{eq:rel_ys_limit}
\end{equation}
we see tat the $y^V$ coordinate cannot change sign because  $\left|  \frac{ e p_U q_V}{32 \Lambda ^2 M^2} \right| \ll 1$, so there exists a horizon as in GR. 

Finally, exchanging the role between left and right particle, and  $U$ and $V$ coordinate, the emission coordinates turn into:
\begin{align}
y^U\,=&\,-\frac{e \epsilon}{\xi_V}\left(1+\frac{3 p_V q_V \xi_V^2-e (p_V q_U+2 q_V (p_U+q_U))}{32 \Lambda ^2 M^2}\right)-\frac{e \xi_V p_V q_V}{32 M^2 \Lambda^2 } \,,\nonumber\\
y^V\,= &\, \xi_U\left(1 + \frac{p_V q_V \xi_V^2-e p_U q_V -2 e (p_V+q_V) q_U}{32 M^2 \Lambda^2}\right)\nonumber \\
&+\frac{\epsilon  \left(-2 e \xi_V^2 (7 p_V q_U+5 p_U q_V+8 q_V q_U)+3 p_V q_V \xi_V^4+2 e^2 p_U q_U\right)}{64 \Lambda ^2 M^2 \xi_V}\,.
\label{eq:rel_ysuv}
\end{align}
\begin{align}
z^U\,= &\,-\frac{e \epsilon}{\xi_V}\left(1+\frac{(p_V+2 q_V) \left(3 p_V \xi_V^2-e p_U\right)}{32 \Lambda ^2 M^2}\right)-\frac{e \xi_V p_V(p_V+2 q_V)}{32 M^2 \Lambda^2 }\,,\nonumber\\ z^V\,= &\, \xi_V\left(1 + \frac{p_V \left((p_V+2q_V)\xi^2_V-e(p_U+2q_U)\right)}{32 M^2 \Lambda^2}\right)\nonumber\\
&+\frac{\epsilon  \left(-4 e \xi_V^2 (2 p_V p_U+3 p_V q_U+p_U q_V)+3 p_V \xi_V^4 (p_V+2 q_V)+2 e^2 p_U (p_U+2 q_U)\right)}{64 \Lambda ^2 M^2 \xi_V}\,.
\label{eq:rel_zsuv}
\end{align}
Imposing $p_V = q_U =0$, the $z$ coordinates tend to:
\begin{equation}
z^U = -\frac{e \epsilon}{\xi_V}\left(1- \frac{e q_V  p_U }{16 \Lambda ^2 M^2}\right) \,,
\qquad
 z^V = \xi_V  +\frac{\epsilon  \left( e^2  p_U^2 -2 e \xi_V^2  p_U q_V \right)}{32 \Lambda ^2 M^2 \xi_V}\,.
\label{eq:rel_zsuv_limit}
\end{equation}
Since $\left| \frac{e q_V  p_U }{16 \Lambda ^2 M^2}
 \right| \ll 1$, the correction terms cannot hope to change the sign of the $z^U$ coordinate. The Schwarzschild horizon of GR is preserved again.

\section{Conclusions}
\label{sec:conclusions}

In this paper we applied the framework introduced in~\cite{Relancio:2020zok,Pfeifer:2021tas} to two specific background spacetimes, namely de Sitter and Schwarzschild. Unlike Minkowski spacetime, these geometries are characterized by a nontrivial causal structure, such that there are observers (assumed to travel along timelike curves) that cannot access the whole of spacetime through the exchange of lightlike signals. To each observer, one can associate a past causal cone of spacetime events that are in causal connection with them, and the region outside of this cone cannot communicate with them. This is one of the most important distinctive features of GR as a physical theory, and its consequences are far-reaching, for example for cosmology, astrophysics, and black hole theory.
A theory like the one developed in~\cite{Relancio:2020zok,Pfeifer:2021tas}, which modifies in a controlled way the notion of locality, but reduces in a certain limit to GR, has the potential to modify the causal structure of GR in a dramatic, potentially catastrophic way.  The notion of horizon and causal diamond might be completely destroyed, if the theory allows to send a signal from any point of spacetime to any other point, by using a sufficiently high-energy particle as messenger. Currently, we are unable to formulate a theorem that is universally applicable to all background spacetimes or deformed kinematics models. However, in order to address this concern, we have begun our investigation by selecting two widely recognized examples of deformed kinematics, and two spacetimes with horizons, and checked whether the nonlocal phenomena predicted by the models destroy the notion of horizon. Interestingly, this is not the case, at least at first order in a perturbative expansion in powers of the high-energy deformation scale.  The notion of horizon seems robust under the type of deformations of the kinematics that the model allows. Thus, this establishes a starting point for future investigations of the fate of GR's causal structure in theories of RL.

\section*{Acknowledgments}
This work has been partially supported by Agencia Estatal de Investigaci\'on (Spain) under grant PID2019-106802GB-I00/AEI/10.13039/501100011033, by the Regional Government of Castilla y Le\'on (Junta de Castilla y Le\'on, Spain), and by the Spanish Ministry of Science and Innovation MICIN and the European Union NextGenerationEU (PRTR C17.I1).
The authors would also like to thank support from the COST Action CA18108.

\end{document}